\documentclass[final,journal,letterpaper,oneside,twocolumn]{IEEEtran}

\usepackage[noadjust]{cite}
\usepackage[pdftex]{graphicx}
\usepackage{array}
\usepackage{amsmath}

\newtheorem{theorem}{Theorem} 
\newtheorem{lemma}[theorem]{Lemma} 
\newtheorem{definition}[theorem]{Definition} 
\newtheorem{corollary}[theorem]{Corollary} 
\newtheorem{example}[theorem]{Example} 

\newtheorem{reptheoreminner}{Theorem}
\newenvironment{reptheorem}[1]{%
  \renewcommand\thereptheoreminner{#1}%
  \reptheoreminner
}{\endreptheoreminner}

\DeclareRobustCommand{\stirling}{\genfrac\{\}{0pt}{}} 
\newcommand{\A}{\mathbf{A}}

\begin{document}
\title{Information Density in Multi-Layer Resistive Memories}%
\author{Susanna~E.~Rumsey,~\IEEEmembership{Graduate Student Member, IEEE}, Stark~C.~Draper,~\IEEEmembership{Senior Member, IEEE}, and Frank~R.~Kschischang,~\IEEEmembership{Fellow, IEEE}%
\thanks{The results of this paper were presented in part at the \textit{Canadian Workshop on Information Theory}, June 2019 \cite{cwit2019} and in Susanna~Rumsey's Master's thesis~\cite{rumsey2019}.\\
This work was supported in part by Discovery Research Grants from the Natural Sciences and Engineering Research Council of Canada (NSERC).\\
Susanna E.~Rumsey, Stark C.~Draper, and Frank R.~Kschischang are with the Department of Electrical and Computer Engineering, University of Toronto, Toronto, ON M5S 3G4, Canada (e-mail: s.rumsey@mail.utoronto.ca; stark.draper@utoronto.ca, frank@ece.utoronto.ca).
}
}

%\markboth{IEEE Transactions on Information Theory,~Vol.~X, No.~X, ~XXX~2020}{Rumsey \MakeLowercase{\textit{et al.}}: Information Density in Multi-Layer Resistive Memories}

%\IEEEpubid{0000--0000/00\$00.00~\copyright~2020 IEEE}

\maketitle

\begin{abstract}
Resistive memories store information in a crossbar arrangement of two-terminal devices that can be programmed to patterns of high or low resistance.  While extremely compact, this technology suffers from the ``sneak-path'' problem: certain information patterns cannot be recovered, as multiple low resistances in parallel make a high resistance indistinguishable from a low resistance.  In this paper, a multi-layer device is considered, and the number of bits it can store is derived exactly and asymptotic bounds are developed.  The information density of a series of isolated arrays with extreme aspect ratios is derived in the single- and multi-layer cases with and without peripheral selection circuitry.  This density is shown to be non-zero in the limit, unlike that of the arrays with moderate aspect ratios previously considered.  A simple encoding scheme that achieves capacity asymptotically is presented.
\end{abstract}

\begin{IEEEkeywords}
Memristors, combinatorial mathematics, resistance, memory architecture, information theory.
\end{IEEEkeywords}

\section{Introduction}

\IEEEPARstart{C}{ompact} non-volatile data storage systems are a key component of modern computer systems.  With advances in materials science and slowing of the shrinking of the transistor technologies that underlie modern solid-state storage, researchers are increasingly exploring alternative storage technologies.  One such technology is the resistive memory array, which has received much attention recently through developments in memristor technology.

Various materials have been considered for resistive memory arrays, including nanowires~\cite{sotiriadis2006} and, more commonly, memristors~(see \cite{pan2014} for a survey).  Memristors are passive two-terminal circuit elements whose resistance can be changed when a sufficiently extreme voltage is applied across the terminals.  The device maintains its new resistance after the applied voltage is removed, and will function as a resistor when less extreme voltages are applied.  Most commonly and most simply, the memristor will change fairly rapidly between a high-resistance state and a low-resistance state, making it a binary device for practical purposes.  The memristor was first proposed as a theoretical device by Chua in 1971~\cite{chua1971}.  However, it was not until 2008 that a feasible method for fabricating such a device was suggested~\cite{strukov2008}.  Since then, many memristor technologies have been explored (see~\cite{pan2014} for an extensive review).  

\begin{figure}[!t]
	\centering
    \includegraphics[scale=0.75]{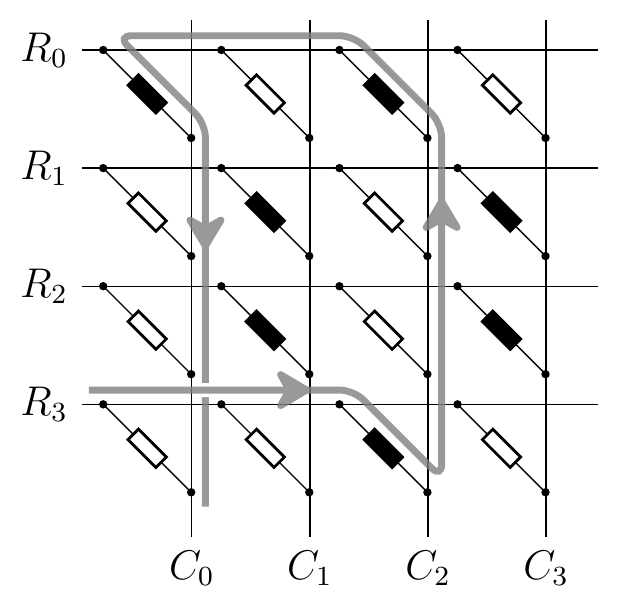}
\caption{A circuit diagram for a sample array, after~\cite{cassuto2016}.  Filled resistors indicate low resistance, and blank resistors indicate high resistance. The sneak path at $(R_3, C_0)$ is drawn explicitly.}
\label{fig:memr_ex}
\end{figure}

In memory storage systems, memristors are typically organized in crossbar fashion, with resistive material separating some number $n_0$ of parallel rows of wires from some number $n_1$ of parallel columns of wires.  A memristor is placed at each row-column intersection, as shown in Figure~\ref{fig:memr_ex}.  With this design we have a single layer of memristors in a rectangle of $n_0$ devices by $n_1$.  The memristors can be programmed to values in a set of resistances, typically a binary set $\{H, L\}$, where $H$ is a high value, and $L$ is a low value.  The ideal case is $H\to\infty$ and $L\to 0$.  We will refer to this architecture more generally as a ``resistive array,'' since it has also been considered in other (non-memristor) resistive memories~\cite{sotiriadis2006}.  This architecture can be generalized to an $\ell$-layer version, with alternating layers of wires of dimensions $n_0 \times n_1\times \dots \times n_\ell$ connected by resistive elements.  (See, e.g.,~\cite{shi2020} for a review of current work on this architecture.)

\IEEEpubidadjcol

The crossbar architecture has many benefits.  First, each memory element has two terminals, which simplifies the device architecture compared to typical transistor-based architectures such as flash.  This results in improvements in density and, depending on the encoding method, possible simplifications in reading and writing schemes.  Second, for manufacturing technology working at a particular technology node, it is possible to make memristors smaller (i.e., with dimensions closer to the node's feature length) without encountering many of the problem, such as reliability and leakage, that occur when transistors for data storage are fabricated at the same scale~\cite{pan2014}.  

Despite these advantages, there is a major problem with the crossbar architecture.  In a large number of patterns of high and low resistances, low-resistance paths known as ``sneak paths" are present in parallel with a single high resistance, with the currents passing through these paths known as ``sneak currents''.  If data is written into such a pattern, this characteristic can cause ambiguity when trying to read an individual resistance.

As an illustrative example of sneak paths, consider the array diagrammed in Figure~\ref{fig:memr_ex}.  Suppose we want to determine the resistance at position $(R_3,C_0)$.  In the example, this device is in a high-resistance state, indicated by an unfilled resistor symbol.  A natural approach to identifying a resistance is to apply a voltage to the wire in row 3 and, through some sensing device, measure the current flowing out of the wire in column~0.  This measurement is clearly a function of resistance at resistor $(R_3, C_0)$.  In this example, however, a low resistance path through resistors $(R_0, C_0)$, $(R_0, C_2)$, and $(R_3, C_2)$ runs parallel to the high resistance at position~$(R_3, C_0)$.  Thus  we measure a much smaller resistance, most likely much closer to the low resistance value, which is typically very small.  Effectively, the high resistance is misread as a low resistance.

The sneak-path problem can be addressed in several ways.  Several sensing approaches have been developed, in which the resistance is estimated by taking several measurements in the circuit~\cite{naous2014, zidan2014}, which have the disadvantage of requiring more circuit measurements per bit read.  Electronic approaches include isolating each resistance with a selection device (e.g., a diode or MOS transistor)~\cite{park2010, shi2020}.  Another method adds peripheral circuitry for redistributing sneak currents \cite{berdan2015}.  These methods have the disadvantage of increasing the size of the device, although progress has been made with placing selection diodes on top of the resistive elements, sometimes at the cost of increased power consumption~\cite{shi2020}.

In this paper, we treat this as a problem in information theory.  Cassuto \textit{et al.}'s work~\cite{cassuto2016} shows that an array will contain sneak paths if and only if there exist some coordinates $(R_x, C_x)$, $(R_x, C_y)$, $(R_y, C_y)$ and $(R_y, C_x)$ (i.e., the corners of some rectangle) exactly three of which indicate low resistance.  Because of this, an array without sneak-paths is referred to variously as an ``isolated zero rectangle free array''~\cite{cassuto2016} or a ``$\Gamma$-avoiding matrix''~\cite{ju2012}, and the problem may be considered a problem in constrained coding. Cassuto \textit{et al.}~propose an encoding scheme without sneak paths that approaches capacity asymptotically.  Their paper in turn has mathematical roots in Sotiriadis' work on nanoscale crossbar devices~\cite{sotiriadis2006}, which derives exact and asymptotic expressions for the capacity.  Similar derivations are made independently in the combinatorics literature~\cite{ju2012}, generalizing the underlying mathematical problems, and deriving some related generating functions.

Building on this prior work, we will take an information-theoretic approach here.  Given a resistive array, we want to develop encoding and decoding schemes that allow us to map a message into a pattern of high and low resistances in the array which may be recovered through a suitable series of measurements.  The present work makes three main contributions:
\begin{enumerate}
\item We derive exact and asymptotic expressions for the capacity of multi-layer devices: the logarithm of the number of patterns $T_\ell(n_0, \dots, n_\ell)$ that can be stored in such a device.  

\item We derive expressions for the information density of single- and multi-layer devices, and demonstrate that a tiled series of isolated arrays with extreme aspect ratios has an improved information density.  

\item We present a simple encoding scheme that achieves capacity asymptotically for such a series of arrays.
\end{enumerate}

The remainder of this paper is divided into five sections.  Section II provides a  mathematical background.  Section III provides a brief overview of existing capacity derivations in single-layer devices, and derives an analogous expression for multi-layer devices.  Section IV derives an asymptotic expression for this capacity.  Section V motivates the use of information density as an important metric for resistive arrays, as distinct from the capacity metric of~\cite{cassuto2016}, and provides an analysis of the optimal size of an isolated array to tile.  Section VI proposes encoding schemes in the single- and multi-layer cases that approach capacity asymptotically.

\section{Mathematical Background}

We will use a graph-theoretical model to describe resistive array states in the idealized model where $H\to \infty$ and \hbox{$L \to 0$}.  Recall that an undirected bipartite graph $G$, denoted $G = (V, W, E)$ consists of two disjoint sets of nodes $V$ and $W$, and a set of edges $E$ of the form $\{v, w\}$ where $v \in V$ and $w \in W$.  If $\{v, w\} \in E$, we say that $v$ and $w$ are neighbors.

\subsection{Single-layer Devices}

We will first consider single-layer devices.  The state of a resistive array with $n_0$ wires on one edge and $n_1$ on the other maps naturally to a matrix representation as follows, equivalent to the notion of a ``configuration'' in~\cite{sotiriadis2006}.

\begin{definition}[Array state matrix]
The array state matrix for a single-layer $n_0 \times n_1$ resistive array is the matrix $\A \in \{0, 1\}^{n_0 \times n_1}$ with elements $\A_{ij}$ satisfying \begin{displaymath}
\A_{ij} = \begin{cases}
0, & \text{if } \mathcal{R}(i, j) = H;\\
1, & \text{if }\mathcal{R}(i, j) = L,
\end{cases}
\end{displaymath} where $\mathcal{R}(i, j)$ is the resistance of the $(i, j)$th element of the array, with $i \in \{0, \dots, n_0 - 1\}$, $j \in \{0, \dots, n_1 - 1\}$.
\end{definition}

Note that in graph-theoretical terms $\A$ is the biadjacency matrix for the sets of row and column wires (e.g., the sets of $R_i$s and $C_i$s in Figure~\ref{fig:memr_ex}).  Conceptually, it is often more useful to map the matrix representation of a resistive array to a graph that represents the connectedness of the array.  We are interested in the following class of graphs that represent how the layers of wires are connected through any low-resistance path.

\begin{definition}[Connectedness graph]
A connectedness graph is a bipartite graph with the property that if two vertices have a common neighbor, they have all their neighbors in common.
\end{definition}

This means that in a connectedness graph $G = (V, W, E)$, if there is any path connecting $v \in V$ and $w \in W$, then $\{v, w\} \in E$.  Such a graph is said to represent an array state matrix $\A$ of an $n_0\times n_1$ resistive array if $V = \{v_0, \dots, v_{n_0 - 1}\}$, $W = \{w_0, \dots, w_{n_1 - 1}\}$, $\A_{ij} = 1$ implies $\{v_i, w_j\} \in E$, and $G$ has no edges that are not necessary to meeting these conditions.  This means that we can uniquely determine whether any possible $\{v_i, w_j\}$ is an element of $E$.  Note that $\{v_i, w_j\} \in E$ does not imply $\A_{ij} = 1$; rather $\{v_i, w_j\} \in E$ implies that there is a path between $v_i$ and $w_j$.  The unique connectedness graph representing the array state matrix $\A$ is denoted as $G(\A)$.\

The connected components of connectedness graphs are always complete bipartite subgraphs (``bicliques'').  These correspond to collections of wires that are connected to each other through low-resistance paths.  If we use a connectedness graph to represent an array state matrix, we cannot distinguish between direct connections and sneak paths; this the connectedness graph reflects the reality of making measurements on the array.

\begin{figure}[!t]
	\centering
    \includegraphics[scale=0.75]{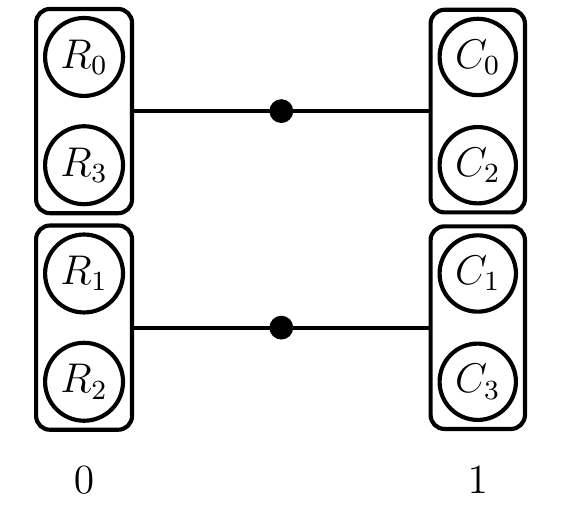}
    
\caption{The dot-graph representation for the circuit in Figure~\ref{fig:memr_ex}.  Wires have been re-ordered for convenience of representation.  Note that wires $R_4$ and $C_1$ belong to the same connected component since they are connected by a sneak path.}
\label{fig:memr_graph}
\end{figure}

Since connectedness graphs always decompose into disjoint complete bipartite subgraphs, we introduce a simplified ``dot-graph'' representation.  Figure~\ref{fig:memr_graph} gives the dot-graph representation for the array of  Figure~\ref{fig:memr_ex}.  The dot-graph representation is based on the bipartite graph representation, in which the two parts of the graph (here labeled `0' and `1') are the sets of row and column wires.  Instead of indicating connections between individual vertices, we indicate connections between distinct \emph{sets} of nodes on each side with an edge with a solid dot.  Thus we have one dotted edge per connected component of the graph.  In this example, the upper dot connection indicates that wires $R_0$, $R_3$, $C_0$, and $C_2$ are all connected, even though in Figure~\ref{fig:memr_ex}, $R_3$ and $C_0$ are connected only through a sneak-path.  As we will see, distinct dot-graphs correspond to distinguishable array state matrices.  This will lead to a natural way to enumerate all distinguishable array state matrices which will yield the storage capacity of the array.

In order to discuss distinguishable array state matrices, we will need the following definitions.

\begin{definition}[Measurement]
A measurement between two disjoint wire sets $\Sigma_V \subseteq V$ and $\Sigma_W \subseteq W$ in a single-layer device with connectedness graph $G = (V, W, E)$ is denoted $\mathcal{M}_G(\Sigma_V, \Sigma_W)$ and is defined by \begin{displaymath}
\mathcal{M}_G(\Sigma_V, \Sigma_W) = \begin{cases}
1 & \text{if some } \{\sigma_V, \sigma_W\} \in \Sigma_V\times \Sigma_W \\
 &\text{satisfies }\{\sigma_V, \sigma_W\} \in E\\
0 & \text{otherwise.}
\end{cases}
\end{displaymath}
\end{definition}

Thus a measurement between two sets indicates whether there is a connection between any pair of their elements.  Physically, this corresponds to applying a voltage to the wires corresponding to the elements of $\Sigma_V$ and measuring whether this causes a current to flow through any of the wires corresponding to $\Sigma_W$.

\begin{definition}[Distinguishable array matrices]
Two array matrices $\A$ and $\A^\prime$ are distinguishable if $G(\A) \ne G(\A^\prime)$.
\end{definition}

Equivalently, two array matrices are distinguishable if there is some array measurement that yields different measurements when applied to each of the two arrays.  For example, if two wires $v_i \in V$ and $w_j \in W$ are in the same connected component in $G(\A)$ and different connected components in $G(\A^\prime)$, we will have $\mathcal{M}_{G(\A)}(\{v_i\}, \{w_j\}) = 1$ and $\mathcal{M}_{G(\A^\prime)}(\{v_i\}, \{w_j\}) = 0$, differentiating $\A$ from $\A^\prime$.

\subsection{Multi-layer Devices}

One of the attractive features of resistive memory arrays is the fact that they are composed of two-terminal devices.  This reduces the number of connectors required, and means that all connectors can be placed at array edges.  One may also stack multiple layers of arrays vertically, producing a very dense device.  In such a device, we alternate layers of wires and layers of resistive material, so that each layer of resistive material and the adjacent two wire layers locally have the same architecture as a single-layer device.  We call these larger devices \emph{multi-layer arrays}.  See~\cite{li2017} and~\cite{shi2020} for electronics and materials science perspectives, and some further illustrations of device architectures.  When counting layers, we will count the number of resistive layers (not the number of wire layers).  Thus, the devices considered so far have been single-layer devices (with two wire layers).

\begin{figure}[!t] 
\centering 
\includegraphics[scale=0.75]{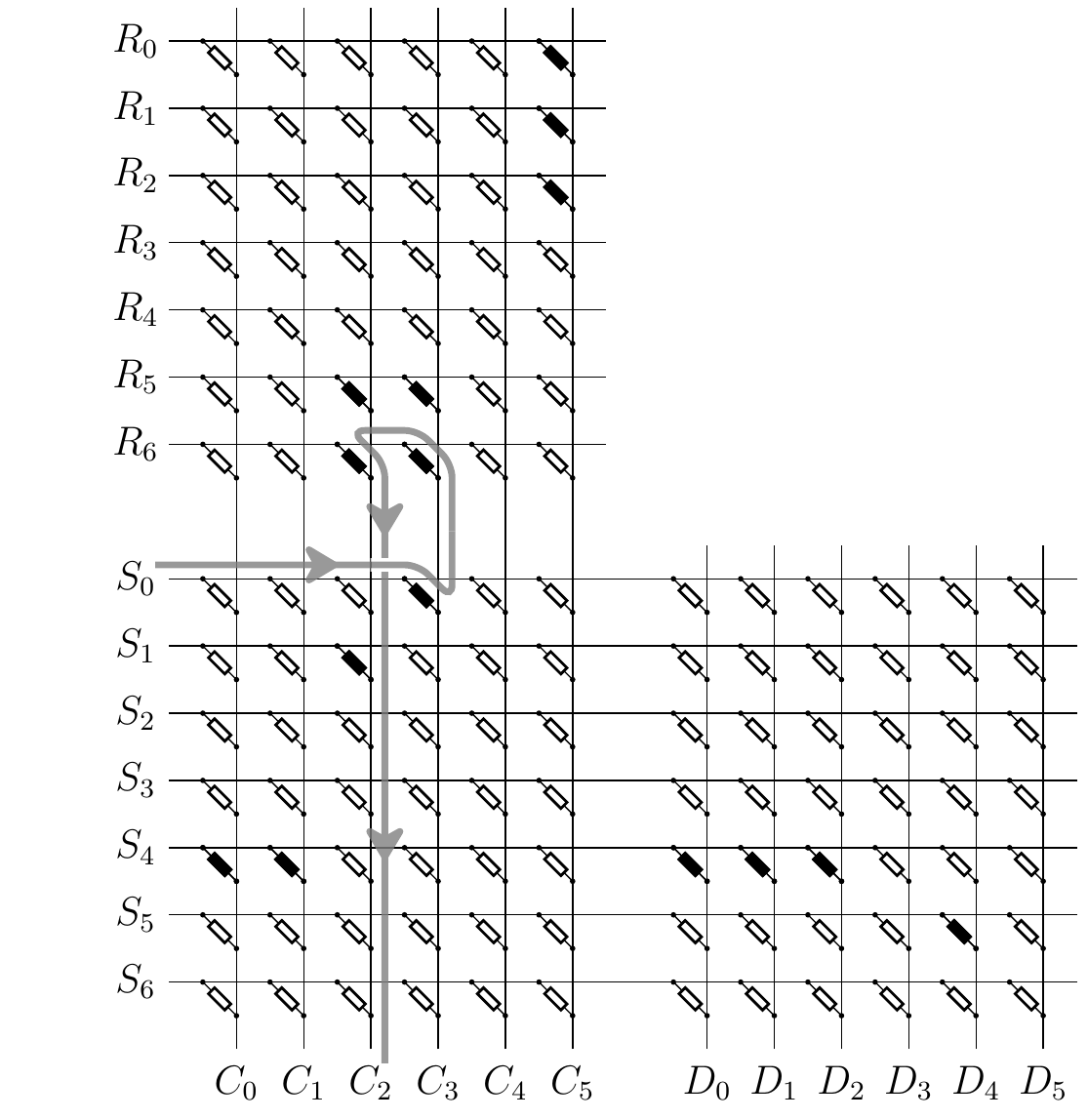} 
 \caption{An example circuit diagram for a three-layer device.  Note the multi-layer sneak path at $(S_0, C_2)$} \label{fig:basicCircuitMulti} 
\end{figure}

An example of a circuit diagram for a three-layer device is given in Figure~\ref{fig:basicCircuitMulti}.  In this example, the top layer consists of memristive devices connecting the $R_i$ and $C_i$ wires.  The middle layer consists of devices connecting the $C_i$ and $S_i$ wires.  The bottom layer consists of devices connecting the $S_i$ and $D_i$ wires.  This means that, for instance, devices $(R_0, C_0)$, $(S_0, C_0)$, and $(S_0, D_0)$ all lie in the same vertical column in the layered device.  (We have expanded the circuit diagram somewhat to make the connections easier to see, but it should be remembered that these components do not all lie in the same plane.)

This circuit also illustrates how a multi-layer device may have sneak paths across two or more layers.  For instance, there is a sneak-path at $(S_0, C_2)$ despite there being no sneak-paths in the local $(S_i, C_i)$ layer.  

We now define the equivalent of the array state matrix for the multi-layer devices.  We will let $n_i$ denote the number of wires in wire layer $i$.

\begin{definition}[Array state matrix vector]
The array state matrix vector for an $\ell$-layer resistive $n_0 \times \cdots \times n_\ell$ device is a vector of matrices $\mathcal{A} = (\A^0, \dots \A^{\ell - 1})$ where $\A^k \in \{0, 1\}^{n_{k - 1}\times n_k}$ has elements $\A^k_{ij}$ satisfying \begin{displaymath}
\A^k_{ij} = \begin{cases}
0 & \text{if }\mathcal{R}^k(i, j) = H;\\
1 & \text{if }\mathcal{R}^k(i, j) = L,
\end{cases}
\end{displaymath}
where $\mathcal{R}^k(i, j)$ is the resistance of the $(i, j)$th element in the $k$th layer.
\end{definition}

Note that this is just the vector of array state matrices for the individual layers of the device.  For instance, the three-layer circuit in Figure~\ref{fig:basicCircuitMulti} has $\mathcal{A} = (\A^0, \A^1, \A^2)$, where $\A^0$ is the array state matrix for the $R_i$s and $C_i$s, $\A^1$ is the array state matrix for the $S_i$s and $C_i$s, and $\A^2$ is the array state matrix for the $S_i$s and $D_i$s.  We now define the multi-layer generalization of the connectedness graph as follows.

\begin{definition}[$\ell$-layer connectedness graph]
An $\ell$-layer connectedness graph is an $(\ell + 1)$-partite graph \begin{displaymath}
G = (V_0,  V_1, \dots, V_\ell, E_1, \dots, E_\ell)
\end{displaymath} where $E_i \subseteq\{(v_{i - 1}, v_i)|v_{i - 1}\in V_{i - 1}, v_i \in V_i\}$, with the property for $i = 1, \dots, \ell$ that if two vertices in $V_i$ have a neighbor in common, they have all their neighbors in common.
\end{definition}

Such an $\ell$-layer connectedness graph is said to represent $(\A^0, \dots, \A^{\ell - 1})$, the array state matrix vector  of a resistive array with $n_i$ wires in wire layer $i$, $i = 0, \dots, \ell$, if $V_i = \{v_{i,0},\dots, v_{i,n_i - 1}\}$ and $\A_{ij}^k = 1$  implies $(v_{k - 1, i}, v_{k, j}) \in E_k$ and $G$ includes only the edges necessary to meet this condition.  Note that $(v_{k - 1,i}, v_{k, j}) \in E_k$ does not imply $\A_{ij} = 1$, rather $(v_{k - 1,i}, v_{k, j}) \in E_k$ implies that there is a path between $v_{k - 1, i}$ and $v_{k, j}$.  The unique connectedness graph representing the array state matrix vector $(\A^0, \dots, \A^{\ell - 1})$ is denoted $G(\A^0, \dots, \A^{\ell - 1})$.

The constrained coding problem for the single-layer case extends naturally to the $\ell$-layer case.  We may mimic the layout of the circuit diagram in Figure~\ref{fig:basicCircuitMulti} by arranging the elements of the array state matrix vector in a ``staircase'' shape of the following form: 
\begin{equation}\label{eq:staircase}
\begin{matrix}
\A^0 \\
(\A^1)^\top & \A^2\\
& (\A^3)^\top\\
& & \ddots\\
& & & (\A^{\ell - 3})^\top &\A^{\ell - 2}\\
& & & &(\A^{\ell - 1})^\top\\
\end{matrix}
\end{equation} where in this case $\ell$ is even and ${}^\top$ indicates a matrix transpose.  The single-layer isolated zero rectangle constraint must apply to the ``staircase'' structure in~(\ref{eq:staircase}).  Equivalently, it must apply to each concatenated matrix $[(\A^i)^\top |\A^{i + 1}]$ for $0 \le i < \ell - 1$.  These concatenated matrices and their transposes correspond to each of the horizontal and vertical ``steps'' in the staircase structure.  This constraint is a result of the fact that a sneak-path across three or more wire layers must include a sneak-path across three wire layers, which is contained in some such concatenated matrix.

\begin{figure}[!t]
\centering
\includegraphics[scale=0.75]{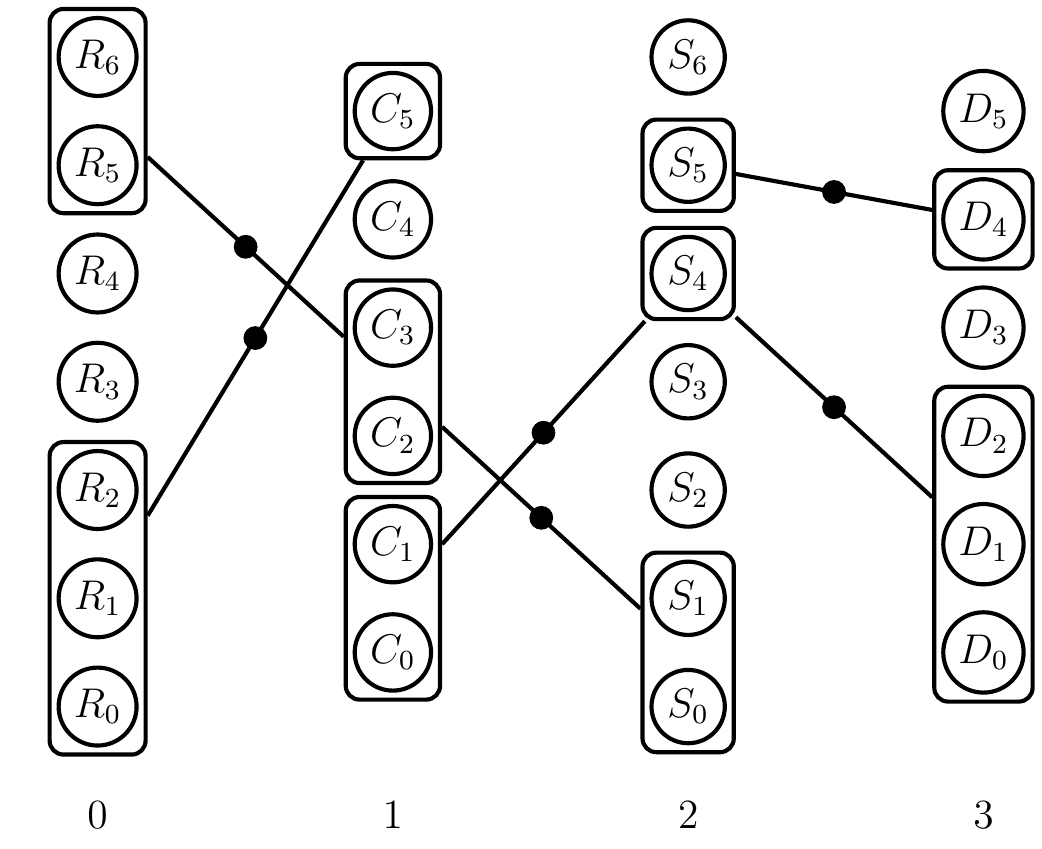}
\caption{An example of the dot-graph representation for the three-layer device from Figure~\ref{fig:basicCircuitMulti} with wire layers of sizes $n_0 = 7$, $n_1 = 6$, $n_2 = 7$, $n_3 = 6$.}
\label{fig:mult_ex}
\end{figure}

The dot-graph representation extends naturally to represent this architecture.  The graph for the circuit in Figure~\ref{fig:basicCircuitMulti} is shown in Figure~\ref{fig:mult_ex}.  As before, a dotted edge between two sets of vertices in this representation corresponds to the complete collection of edges between the two sets of wires in adjacent wire layers (i.e., the biclique).

\section{Counting Sneak-Path-Free Arrays}

The capacity of a single-layer device has been derived in several places in the literature~\cite{sotiriadis2006,ju2012}.  We present another proof here that eases the generalization to the multi-layer case.

\subsection{Single-layer Device Capacity}

To determine the number of distinguishable resistive array state matrices, we count the number of distinct connectedness graphs.  This count is a function of the array dimensions $n_0$ and $n_1$ and we denote it $T_1(n_0, n_1)$.  Since this involves partitioning the input and output wires, it will be useful to recall the following definition, which is applicable when counting partitions.

\begin{definition}[Stirling number of the second kind]\label{def:stirling}
The Stirling number of the second kind, denoted $\stirling{n}{k}$, is the number of ways in which a set of $n$ items may be partitioned into $k$ (non-empty) sets.  For $n > 0$ and $k > 0$ it is equal to \begin{equation}\label{eq:stirling}
\stirling{n}{k} = \frac1{(k - 1)!}\sum_{i = 0}^{k - 1} (-1)^i \binom{k - 1}{i}(k -i)^{n - 1},
\end{equation} a reindexed version of the form given in~\cite[Ch.~1.9]{stanley2012}, which will be useful in subsequent derivations.  Note that $\stirling{0}{0} = 1$, and $\stirling{n}{0} = \stirling{0}{k} = 0$ when $n > 0$ and $k > 0$.
\end{definition}

Both Sotiriadis and later Ju and Seo derive the value of $T_1(n_0, n_1)$ using various counting arguments~\cite{sotiriadis2006, ju2012}.  The proof we present here is similar to the latter, but stated in a form that is more useful to the current application.

\begin{theorem}[{\cite[Thm.~1]{sotiriadis2006}}{\cite[Thm.~3.1]{ju2012}}]\label{thm:T2}
There are
\begin{equation}\label{eq:T2}
T_1(n_0, n_1) = \sum_{k = 0}^{\min(n_0, n_1)}
      \stirling{n_0+1}{k+1}\stirling{n_1+1}{k+1}k!
\end{equation}
distinct $n_0 \times n_1$ connectedness graphs.
\end{theorem}

\begin{IEEEproof}
Consider the layer of $n_0$ wires.  Label these wires $0, \dots, n_0 - 1$ and partition these labeled wires, plus the symbol $*$, into $k + 1$ non-empty subsets.  There are $\stirling{n_0 + 1}{k + 1}$ ways to do this.  Repeat for the layer of $n_1$ wires.  In each case, one subset contains the symbol $*$.  Any wires in these subsets will remain unconnected.  Connect the remaining $k$ subsets of the layer of $n_0$ wires pairwise with the remaining $k$ subsets of the layer of $n_1$ wires.  This can be done in $k!$ ways.  Summing over $k$ gives the desired final expression for~$T_1$.  
\end{IEEEproof}

Patterns with different values of $k$ are distinct, since $k$ is the number of connected components in the graph.  Note that $T_1(n_0, n_1) = T_1(n_1, n_0)$.  The generating function for $T(n_0, n_1)$ is $\Phi(x, y) = \exp[(e^x - 1)(e^y - 1) + x + y]$~\cite{ju2012} and the sequence $T(n_0, n_0)$ is number A014235 in the On-Line Encyclopedia of Integer Sequences (OEIS)~\cite{OEIS2020}.

\subsection{Single-layer Capacity Asymptotics}

The exact data storage capacity in bits of an $n_0\times n_1$ array is $\log T_1(n_0, n_1)$.\footnote{All logarithms are base 2 unless otherwise specified.} This expression is difficult to manipulate, however, so it is useful to develop an asymptotic scaling for the capacity.  We will consider sequences that are asymptotic in the following sense, following~\cite{sotiriadis2006}.
\begin{definition}\label{def:asymp}
If $\eta_n$ and $\zeta_n$ are positive sequences with $\eta_n, \zeta_n \ne 1$ for sufficiently large $n$, we write $\eta_n \sim \zeta_n$ if $\log(\eta_n)/\log(\zeta_n) \to 1$ as $n\to \infty$.
\end{definition}

An asymptotic approximation of the capacity for large $n_0 + n_1$ is~\cite[Thm.~4]{sotiriadis2006}\begin{equation}
\log T_1(n_0, n_1) \sim (n_0 + n_1)\log(n_0 + n_1).
\label{eq:asymp}
\end{equation}

While this approximation may be useful for describing the limiting behavior of the system, it does not necessarily give useful numerical results, especially for smaller values of $n_0$ and $n_1$.  For instance, numerical results for the $n_0 = n_1$ case suggest that the ratio $2n_0\log(2n_0)/\log T(n_0, n_0)$ is decreasing in $n_0$, but only decreases to approximately 1.45 at $n_0 = 4800$.

\subsection{Multi-layer Device Capacity}

We now turn to the $\ell$-layer case.  Consider an $\ell$-layer device with dimensions $n_0 \times \cdots \times n_\ell$.    We will count the number of distinct connectedness graphs, denoted as $T_\ell(n_0, \dots, n_\ell)$ as follows, beginning with a relevant combinatorial definition.

\begin{definition}[Trinomial coefficient]
The trinomial coefficient, denoted $\binom{n}{i, j, k}$, with $i + j + k = n$, defined as \begin{displaymath}
\binom{n}{i, j, k} = \frac{n!}{i!j!k!}
\end{displaymath} is the number of ways in which $n$ objects may be partitioned into subsets of sizes $i$, $j$, and $k$.
\end{definition}

\begin{theorem}\label{thm:TMulti}
There are
	\begin{multline}\label{eq:numpatts}
	T_\ell(n_0, \dots, n_\ell) =\\
	 \sum_{s_1 = 0}^{\min(n_0, n_1)}\cdots\sum_{s_\ell = 0}^{\min(n_{\ell - 1}, n_\ell)}\prod_{i = 0}^{\ell}\sum_{k_i = \max(s_i, s_{i + 1})}^{\min(n_i, s_i + s_{i + 1})} s_i!\stirling{n_i + 1}{k_i + 1}\\
	 \binom{k_i}{s_i + s_{i + 1} - k_i, k_i - s_{i + 1}, k_i - s_i}
	\end{multline}
distinct $\ell$-layer connectedness graphs for an $ n_0\times \cdots \times n_\ell$ device, where $s_0 = s_{\ell + 1} = 0$.
\end{theorem}

\begin{IEEEproof}
We have $\ell + 1$ wire layers of sizes $n_0, \dots, n_\ell$ respectively.  Consider the layer of $n_0$ wires.  Label them $0, \dots, n_0 - 1$ and partition these wires, plus the symbol $*$, into $k_0 + 1$ non-empty subsets, as in the single-layer case.  There are $\stirling{n_0 + 1}{k_0 + 1}$ possible partitions.  Repeat for each layer, giving $\stirling{n_i + 1}{k_i + 1}$ possible partitions at wire layer $i$.  (These are the partitions that are represented with rounded rectangles in the connectedness graph.)

In wire layer $i$, we connect the $k_i$ partitions that do not include the symbol $*$ to some of the corresponding partitions in layers $i - 1$ and $i + 1$ (when they exist).  We will connect $L_i$ of these to partitions in layer $i - 1$ only, $U_i$ to partitions in layer $i + 1$ only, and the remaining $B_i := k_i - U_i - L_i$ to partitions in both.  There are $\binom{k_i}{B_i, L_i, U_i}$ ways to do this.  This means that there are $s_i := B_i + L_i = B_{i - 1} + U_{i - 1}$ connections between partitions in layer $i - 1$ and partitions in layer $i$, which can be made in $s_i!$ possible ways.  We must therefore sum over all values of 
\begin{IEEEeqnarray*}{lCr}
\IEEEeqnarraymulticol{3}{l}{
s_i!\stirling{n_i +1}{k_i + 1}\binom{k_i}{B_i, L_i, U_i}}\\
\qquad&=& s_i!\stirling{n_i + 1}{k_i + 1}\binom{k_i}{s_i + s_{i + 1} - k_i, k_i - s_{i + 1}, k_i - s_i}.
\end{IEEEeqnarray*}
The bounds on the sums in~(\ref{eq:numpatts}) correspond to the terms that give non-zero Stirling numbers and trinomial coefficients.
\end{IEEEproof}

Since $s_i$ describes the number of disjoint connections between two wire layers (the number of dots in a layer in the dot-graph representation), and $k_i$ describes the number of disjoint components in a wire layer (the number of rectangles in a wire layer), connectedness graphs with different values of $s_i$s or $k_i$s will always be distinct, and represent distinguishable array state matrix vectors.  Note that when $\ell=1$,~(\ref{eq:numpatts}) simplifies to $T_1(n_0, n_1)$ given in~(\ref{eq:T2}).

To illustrate the idea of the proof, consider the device represented by the graph in Figure~\ref{fig:mult_ex}.  It has $n_0 = n_2 = 7$, and $n_1 = n_3 = 6$.  Wire layer~0 is partitioned into $k_0 = 2$ subsets of sizes 2 and 3, with the remaining 2 wires disconnected.  Since this is the first layer, $B_0 = L_0 = 0$, so $U_0 = k_0 = 2$.  There are then $s_1 = k_0 = 2$ connections across the first resistor layer.  These correspond to the two dot-connections in the first layer in the diagram.

In wire layer 1, we have $k_1 = 3$ subsets of wires with one wire disconnected.  Out of these 3 subsets, $L_1 = 1$ is connected to the lower layer only, $B_1 = 1$ is connected to both the lower and upper layer, and $U_1 = 1$ is connected to the upper layer only.  There are $s_1 = B_1 + L_1 = 2$ subsets that connect to the lower layer, and $s_2 = B_1 + U_1 = 2$ subsets connecting to the upper layer.

We partition the upper layers similarly, with $B_2 = L_2 = U_2 = 1$, $L_3 = 2$, and $B_3 = U_3 = 0$.  Equivalently, $k_2 = 3$, $k_3 = 2$, and $s_3 = 2$.

\subsection{Multi-layer Capacity Asymptotics}

To develop an asymptotic expression more amenable to interpretation and to manipulation, we first consider the case where all layers have $n$ wires, analogously to the single-layer derivation in \cite{sotiriadis2006}.  Define $T_\ell(n) := T_\ell(n, \dots, n)$ and \begin{equation}\label{eq:Q}
Q_n(k, a, b) := a!\binom{k}{a+ b - k, k - a, k - b}\stirling{n + 1}{k + 1}.
\end{equation}

Expanding the sums over $k_0$ and $k_\ell$ in~(\ref{eq:numpatts}) gives the alternative form
	\begin{multline}\label{eq:tfunc}
	 T_\ell(n) = \sum_{s_1 = 0}^n\cdots\sum_{s_\ell = 0}^n\stirling{n + 1}{s_1 + 1}\stirling{n + 1}{s_\ell + 1}s_\ell!\\
	\prod_{i = 1}^{\ell - 1} \sum_{k_i = \max(s_i, s_{i + 1})}^{\min(n, s_i + s_{i + 1})}Q_n(k_i, s_i, s_{i + 1}).
	\end{multline} 

We will use the following result from~\cite{sotiriadis2006}.
\begin{lemma}[\cite{sotiriadis2006}, (20)]\label{lem:lb}
Given integers $n$ and $\rho$, with $2 < \rho < n$, define $[n/\rho] := 1 + \lfloor n/\rho\rfloor$.  Then \begin{displaymath}\label{eq:lb}
\stirling{n}{[n/\rho]} > \frac{n!}{[n/\rho]!}\frac1{(\rho!)^{[n/\rho]}}.
\end{displaymath}
\end{lemma}

We obtain a lower bound as follows.

\begin{lemma}\label{lower}
For every integer $\rho$ such that $2 < \rho < n$, there exists a function $L_\ell^\rho(n)$  such that $T_\ell(n) > L_\ell^\rho(n)$ and $L_\ell^\rho(n) \sim n^{(\ell + 1 - 1/\rho)n}$, in the sense of Definition~\ref{def:asymp}.
\end{lemma}

\begin{IEEEproof}
This proof is analogous to the single-layer proof in~\cite{sotiriadis2006}.  

The function $T_\ell(n)$ is increasing in $n$.  Therefore, using~(\ref{eq:tfunc}),
	\begin{IEEEeqnarray*}{rCl}
	T_\ell(n) &>& T_\ell(n - 1)\\
	 &=& \sum_{s_1 = 1}^{n}\cdots\sum_{s_\ell = 1}^{n}\stirling{n}{s_1}\stirling{n}{s_\ell}(s_\ell - 1)!\\
	 &&\;\prod_{i = 1}^{\ell - 1} \sum_{k_i = \max(s_i, s_{i + 1})}^{\min(n, s_i + s_{i + 1} - 1)} Q_{n - 1}(k_i - 1, s_i - 1, s_{i + 1} - 1).
	\end{IEEEeqnarray*}

Take an integer parameter $\rho$ as in Lemma~\ref{lem:lb}.  Lower bound each sum over $s_i$ by the $s_i = [n/\rho]$ term.  This gives
\begin{multline}\label{eq:lb_initial}
 T_\ell(n) > \stirling{n}{[n/\rho]}^2([n/\rho] - 1)!\prod_{i = 1}^{\ell - 1}\sum_{k_i = [n/\rho]}^{\min(n, 2[n/\rho] - 1)}\\
Q_{n - 1}(k_i - 1, [n/\rho] - 1, [n/\rho] - 1).
\end{multline} Similarly, lower bound each sum over $k_i$ by the $k_i = [n/\rho]$ term, and apply Lemma~\ref{lem:lb}, giving 
\begin{IEEEeqnarray*}{rCl}
T_\ell(n) &>& \left([n/\rho] - 1)!\right)^\ell\stirling{n}{[n/\rho]}^{\ell + 1}\\
&>& (([n/\rho] - 1)!)^\ell\left(\frac{n!}{[n/\rho]!}\cdot\frac1{(\rho!)^{[n/\rho]}}\right)^{\ell + 1}\\
&=& \frac{\left(n!/(\rho!)^{[n/\rho]}\right)^{\ell + 1}}{[n/\rho]^\ell[n/\rho]!}\\
&>& \frac{\left(n!/(\rho!)^{\frac{n}{\rho} + 1}\right)^{\ell + 1}}{\left(n/\rho + 1\right)^\ell[n/\rho]!}\\
&:=& L_\ell^\rho(n).
\end{IEEEeqnarray*}

Using the properties of the $\sim$ relationship derived in~\cite{sotiriadis2006}, \begin{displaymath}
L_\ell^{\rho}(n) \sim \frac{(n!)^{\ell + 1}}{[n/\rho]!} \sim \frac{n^{(\ell + 1)n}}{n^{n/\rho}} = n^{(\ell + 1 - 1/\rho)n}
\end{displaymath} as desired.
\end{IEEEproof}

To obtain an upper bound, we leverage the following bound, also found in~\cite{sotiriadis2006}.

\begin{lemma}[\cite{sotiriadis2006}, Lem.~A1]\label{ineq}
For $n \ge m$, \begin{displaymath}
\stirling{n}{m} \le \frac{n!}{m!}(e + 1)^m.
\end{displaymath}
\end{lemma}

We now state and prove the upper bound.

\begin{lemma}\label{upper}
There exists a function $U_\ell(n)$  such that $T_\ell(n) < U_\ell(n)$ and $U_\ell(n) \sim n^{(\ell + 1)n}$.
\end{lemma}

\begin{IEEEproof}
Since $3^k = \sum_{a, b, c}\binom{k}{a, b, c}$, we may write 
	\begin{IEEEeqnarray*}{rClr}
	T_\ell(n) &\le& \sum_{s_1 = 1}^n\cdots\sum_{s_\ell = 1}^n\stirling{n + 1}{s_1 + 1}\stirling{n + 1}{s_\ell + 1}s_\ell!\\
	 && \qquad\prod_{i = 1}^{\ell - 1}\sum_{k_i = \max(s_i, s_{i + 1})}^{\min(n, s_{i - 1} + s_i)}k_i!3^{k_i}\stirling{n + 1}{k_i + 1}\\
 &\le& (n + 1)^{\ell - 2}\prod_{i = 1}^{\ell - 1}\left(\sum_{k_i = 0}^n k_i!3^{k_i}\stirling{n + 1}{k_i + 1}\right)\\
 &&\qquad\left(\sum_{s_1 = 0}^n\stirling{n + 1}{s_1 + 1}\right)\left(\sum_{s_\ell = 0}^n\stirling{n + 1}{s_\ell + 1}s_\ell!\right)
  .
	\end{IEEEeqnarray*}

Applying Lemma~\ref{ineq} gives 
\begin{multline*}
	T_\ell(n) \le (n + 1)!^{\ell + 1}(n + 1)^{\ell - 2}\prod_{i = 1}^{\ell - 1}\left(\sum_{k_i = 0}^n3^{k_i}\frac{(e + 1)^{k_i + 1}}{k_i + 1}\right)\\
	\quad\left(\sum_{s_1 = 0}^n\frac{(e + 1)^{s_1 + 1}}{(s_1 + 1)!}\right)\left(\sum_{s_\ell = 0}^n\frac{(e + 1)^{s_\ell + 1}}{s_\ell + 1}\right).
\end{multline*}

The sum in $s_1$ can be bounded by the Taylor series of an exponential.  Since the other sums are increasing in the variable of summation, each summand can be bounded by the $n$th term.  This gives 
	\begin{IEEEeqnarray*}{rCl}
	 T_\ell(n) &\le& (n + 1)!^{\ell + 1}(n + 1)^{\ell - 2}\left(3^n(e + 1)^{n + 1}\right)^{\ell - 1}\\
	 &&\qquad(e^{e + 1} - 1)(e + 1)^{n + 1}\\
	 &:=& U_\ell(n)\\
	 &\sim& n^{(\ell + 1)n}.
	\end{IEEEeqnarray*}
Therefore, $U_\ell(n)$ has the desired properties.
\end{IEEEproof}

We now derive an asymptotic expression for $T_\ell$.

\begin{theorem}\label{thm:asymp}
The function $T_\ell(n)$ satisfies
\begin{displaymath}
T_\ell(n) \sim n^{(\ell + 1)n}.
\end{displaymath}
\end{theorem}

\begin{IEEEproof}
From Lemmas~\ref{lower} and~\ref{upper}, we know that for any integer $\rho$ such that $2 < \rho < n$, we may bound $T_\ell$ by $L_\ell^\rho(n) < T_\ell(n) < U_\ell(n)$.  This means that \begin{multline}\label{eq:logBounds}
\left(1 - \frac1{(\ell + 1)\rho}\right)\frac{\log(L_\ell^\rho(n))}{\log(n^{(\ell + 1 - 1/\rho)n})} <\\ \frac{\log(T(n))}{\log(n^{(\ell + 1)n})}
 < \frac{\log(U_\ell(n))}{\log(n^{(\ell + 1)n})}.
\end{multline}

By Definition~\ref{def:asymp} and Lemmas~\ref{lower} and~\ref{upper},\begin{displaymath}
\lim_{n\to\infty}\frac{\log(L_\ell^\rho(n))}{\log(n^{(\ell + 1 - 1/\rho)n})} = \lim_{n\to\infty}\frac{\log(U_\ell(n))}{\log(n^{(\ell + 1)n})} = 1.
\end{displaymath}  Applying this result to~(\ref{eq:logBounds}) gives \begin{multline*}
\left(\ell + 1 - \frac1\rho\right) \le \lim_{n\to\infty}\inf\frac{\log(T_\ell(n))}{\log n^n} \\ \le \lim_{n\to\infty}\sup\frac{\log(T_\ell(n))}{\log n^n} \le \ell + 1
\end{multline*}
for every $\rho > 2$.  Taking $\rho \to\infty$ gives $\lim_{n\to\infty}\frac{\log(T_\ell(n))}{\log n^n} = \ell + 1$, so $T_\ell(n) \sim n^{(\ell + 1)n}$.
\end{IEEEproof}

This asymptotic result reveals a drawback of this type of memory with this uniform geometry: the number of bits stored per unit area is \begin{equation}\label{eq:badDensity}
\frac{\log T_\ell(n)}{n^2} \sim \frac{(\ell + 1)n\log n}{n^2} = (\ell + 1)\frac{\log n}{n}
\end{equation} which goes to 0 as $n$ becomes large.

We can make a similar claim for devices whose wire layers differ in size by a constant factor in the limit:

\begin{theorem}
For a multi-layer device with wire layer dimensions $n_0, \dots, n_\ell$,
\begin{equation}
T_\ell(n_0, \dots, n_\ell) \sim \prod_{i = 0}^\ell n_i^{n_i}
\end{equation} when $n_i \to \infty$ for each $i$, and $n_i/n_j \to a_{ij} > 0$ for each pair $(i, j)$ and some constant $a_{ij}$.
\end{theorem}

The proof of this is similar to that of Theorem~\ref{thm:asymp}: for both bounds, we proceed analogously except that bounds on sums over $s_i$ and $k_i$ are taken in terms of $n_i$ instead of $n$.  The requirement that the wire layer aspect ratios stay constant as we take the limit means that our density still goes to zero analogously to~(\ref{eq:badDensity}).

\section{Density Optimization}

Not all array aspect ratios have an information density that goes to zero, which means that tiling isolated arrays of these sizes will make a larger array with density bounded away from zero.  In this section we explore the information density of these arrays, and derive the optimal size for maximizing density.

\subsection{Density of High-Aspect-Ratio Single-Layer Devices}

The approximation to single-layer capacity developed in~(\ref{eq:asymp}) is only applicable when $n_0/n_1$ approaches a finite positive constant as $n_0$ and $n_1$ increase.  We know, for instance, that the density is 1 in the case of a $1\times n_1$ device, since it is impossible to create a sneak path in such a device.  
Building on this observation, we will consider the case where $n_0$ becomes large while $n_1$ remains fixed.  We first derive bounds on $T_1(n_0, n_1)$.

\begin{theorem}
\label{thm:stirlingBoundUpper} If positive integers $n_0$ and $n_1$ satisfy \begin{equation}\label{eq:singleLayerBound}
n_0 \ge \log_{(1 + \frac1{n_1})}\frac{n_1(n_1 + 1)}2
\end{equation}
then
 \begin{displaymath} T_1(n_0, n_1) \le(n_1 + 1)^{n_0 + 1}. \end{displaymath}
\end{theorem} 
\begin{IEEEproof}
See Appendix~\ref{app:singleProof}.
\end{IEEEproof}

\begin{theorem}\label{thm:stirlingBoundLower} 
For positive integers $n_0$ and $n_1$,\begin{displaymath} 
(n_1 + 1)^{n_0} \le T_1(n_0, n_1). 
\end{displaymath} 
\end{theorem} 
\begin{IEEEproof} 
The subset of $n_0\times n_1$ array state matrices having at most a single 1 in any row, of which there are $(n_1 + 1)^{n_0}$, includes only sneak-path-free array state matrices.
\end{IEEEproof}

Taking logarithms of Theorems~\ref{thm:stirlingBoundUpper} and~\ref{thm:stirlingBoundLower} gives the following.

\begin{corollary}\label{corr:bitBounds}
The number of bits \begin{displaymath}
B(n_0, n_1) :=\log T_1(n_0, n_1)
\end{displaymath} that can be stored in an $n_0\times n_1$ array with dimensions satisfying~(\ref{eq:singleLayerBound}) satisfies \begin{equation} \label{eq:singleBounds}
n_0\log(n_1 + 1) \le B(n_0, n_1) \le(n_0 + 1)\log(n_1 + 1). \end{equation} 
\end{corollary}

Dividing~(\ref{eq:singleBounds}) by the array area $n_0n_1$ gives the following corollary.

\begin{corollary}\label{cor:densityBounds} 
The bit density of an $n_0\times n_1$ array with dimensions satisfying~(\ref{eq:singleLayerBound}) satisfies \begin{displaymath} 
\frac{\log(n_1 + 1)}{n_1} \le\frac{B(n_0, n_1)}{n_0n_1} \le\frac{n_0 + 1}{n_0}\frac{\log(n_1 + 1)}{n_1}. 
\end{displaymath} 
\end{corollary}

Taking $n_0\to\infty$ gives the following result.

\begin{corollary}\label{cor:densityConv} 
For $n_1$ fixed and $n_0\to\infty$, the bit density of an $n_0\times n_1$ array converges to $n_1^{-1}\log(n_1 + 1)$ from above. 
\end{corollary}

This density is bounded away from zero for fixed $n_1$.  In particular, it has maximal value 1 at $n_1 = 1$, meaning that we have one bit per resistive cell if we have a $1\times n_0$ array.

We can also derive an asymptotic expression for $T_1$ for fixed $n_1$ and $n_0 \to \infty$.

\begin{theorem}\label{thm:TLimitSingle}
For any fixed $n_1$, \begin{displaymath}
\lim_{n_0\to\infty} \frac{T_1(n_0, n_1)}{(n_1 + 1)^{n_0}} = 1
\end{displaymath} and this limit is approached from above.
\end{theorem}
\begin{IEEEproof}
Using~(\ref{eq:stirling}), we can write \begin{IEEEeqnarray*}{rCl}
T(n_0,n_1) &=& \sum_{i=0}^{n_1} \stirling{n_0+1}{i+1}
 \stirling{n_1+1}{i+1} i!\\
 &=& \sum_{i=0}^{n_1} \stirling{n_1+1}{i+1}\left(\sum_{j=0}^i (-1)^{j} \binom{i}{j} (i+1-j)^{n_0} \right)\\
 &=& \sum_{k=0}^{n_1} \sum_{i=k}^{n_1}  (-1)^{i-k} \stirling{n_1+1}{i+1}
 \binom{i}{k} (k+1)^{n_0} \\
 &=& \sum_{k=0}^{n_1} a(k,n_1) (k+1)^{n_0}
\end{IEEEeqnarray*} where \begin{displaymath}
a(k, n_1) = \sum_{i = k}^{n_1}(-1)^{i - k}\stirling{n_1 + 1}{i + 1}\binom{i}{k}.
\end{displaymath}  Since $a(n_1, n_1) = 1$, $T(n_0, n_1) = (n_1 + 1)^{n_0} + \mathcal{O}(n_1^{n_0})$, so $\lim_{n_0\to\infty} T(n_0, n_1)/(n_0 + 1)^{n_1} = 1$ as desired.
\end{IEEEproof}

\subsection{Tiling Isolated Arrays}

Despite maximizing the limiting density described in Corollary~\ref{cor:densityConv}, a $1\times n_0$ resistive array becomes impractical to use for large values of $n_0$.  A more common solution to the sneak-path problem is to isolate each memory cell with a selection device such as a diode or a transistor (see e.g.,~\cite{park2010} and~\cite{shi2020}).  Using diodes is particularly appealing, since they are two-terminal elements and therefore require less layout and do not need individual switching signals.  The density of such a device does not change with its size or shape, but a non-negligible fraction of the circuit area may be used for the diodes themselves rather than the memory cells, thereby reducing the density of memristive devices (though perhaps not the information density).  It is possible in some cases to stack a small selection device on top of the memristor, which does not require additional area.  This architecture can have increased power consumption, which means it may not be desirable for all applications~\cite{shi2020}.  Another alternative that may increase power consumption involves grounding a large fraction of the wires in one direction~\cite{cassuto2016}.

\begin{figure}[!t] 
\centering 
\includegraphics[width=\columnwidth]{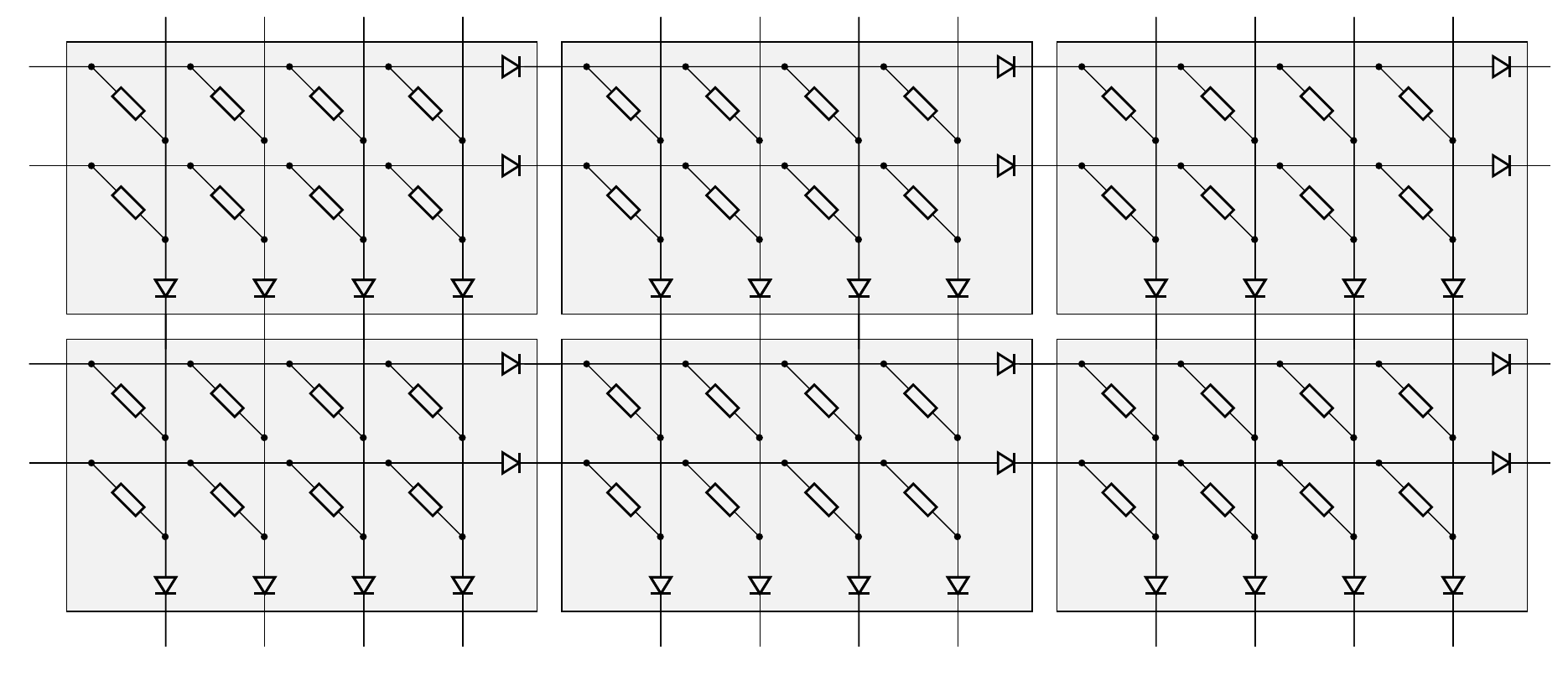} 
\caption{A $2\times 4$ crossbar array with diodes along the edges tiled to make a larger array.}\label{fig:edgeDiode} 
\end{figure}

We propose an alternative approach.  Consider an $n_0\times n_1$ crossbar array with memory cells of unit area.  Let us assume the area of each diode is $\delta$ times the area of a memristive cell.  Place diodes on each of the $n_0 + n_1$ output lines, and tile the resulting isolated array as shown in Figure~\ref{fig:edgeDiode}.  The diodes isolate the arrays, making it impossible for sneak-paths to form.  This architecture will be referred to as an $(n_0, n_1)_\delta$ device.

The resulting array has area $A_\delta^{\textnormal{arr}}(n_0, n_1) := n_0n_1 + 
\delta(n_0 + n_1)$.  We may bound the array density $\rho_\delta^{\textnormal{arr}}(n_0, n_1) := B(n_0, n_1)/A_\delta(n_0, n_1)$ using the same approach as in Corollaries~\ref{cor:densityBounds} and~\ref{cor:densityConv}, giving the following further corollaries to Theorems~\ref{thm:stirlingBoundUpper} and~\ref{thm:stirlingBoundLower}.

\begin{corollary}\label{cor:diodeDensityBounds} 
For an $(n_0, n_1)_\delta$ device, if $n_0$ satisfies~(\ref{eq:singleLayerBound}), then \begin{displaymath} 
\frac{n_0\log(n_1 + 1)}{n_0n_1 + \delta(n_0 + n_1)} \le\rho^{\textnormal{arr}}_\delta(n_0, n_1) \le\frac{(n_0 + 1)\log(n_1 + 1)}{n_0n_1 + \delta(n_0 + n_1)}. 
\end{displaymath} 
\end{corollary}

\begin{corollary}
For $n_1$ fixed and $n_0\to\infty$, the array bit density $\rho^{\textnormal{arr}}_\delta(n_0, n_1)$ of an $(n_0, n_1)_\delta$ device converges to $P^{\textnormal{arr}}_\delta(n_1) := (n_1 + \delta)^{-1}\log(n_1 + 1)$. 
\end{corollary}

\begin{figure}[!t] 
\centering 
\includegraphics[scale=0.75]{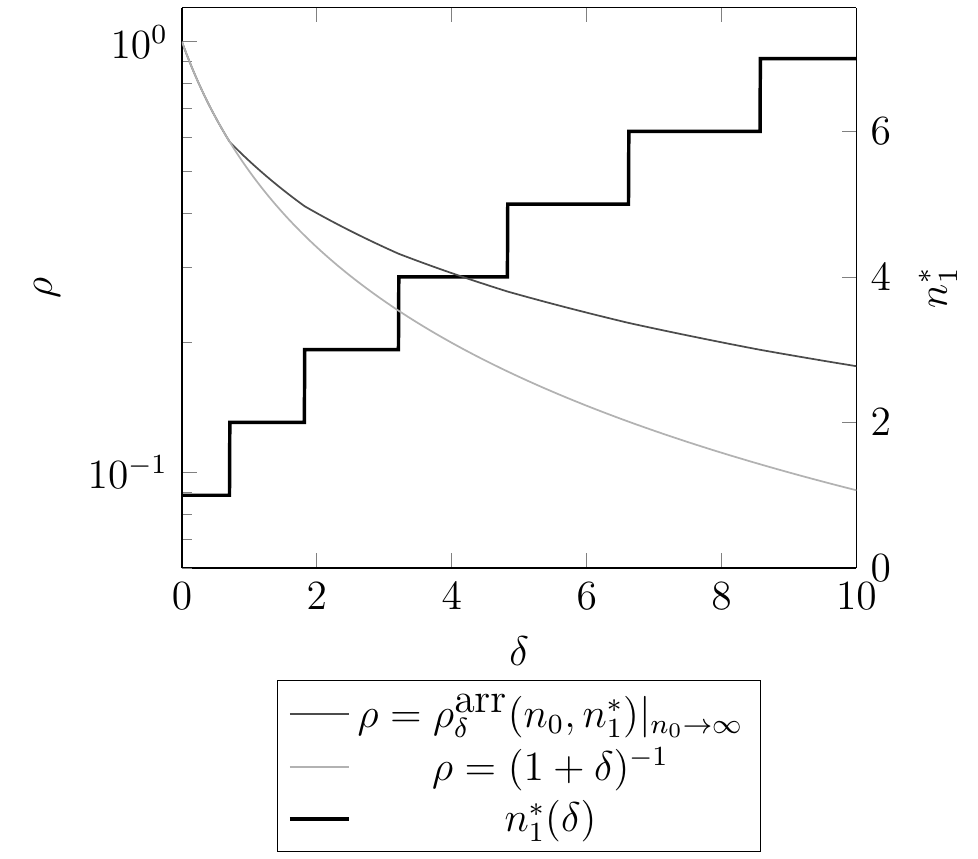}
\caption{Optimal array dimension $n_1^*$ versus diode area factor $\delta$ in the $n_0\to\infty$ regime (right axis) and the resulting bit density for this choice of $n_1$, compared to the na\"\i{ve} series diode approach (left axis).} \label{fig:nStar} 
\end{figure}

While the value of $n_1$ that maximizes $P_\delta^\textrm{arr}(n)$ can be derived using methods from complex analysis, in practice we will usually evaluate it numerically.  The optimal integer value $n_1^*$ is plotted against different values of $\delta$ in Figure~\ref{fig:nStar}.  This figure also shows the bit densities at these choices of $n_1^*$, with the $\rho= (1 + \delta)^{-1}$ curve from the existing architecture with one diode per array unit for reference.  We see that in all cases, the architecture with fewer diodes has a better bit density than the architecture with diodes at all locations.

\begin{figure}[!t] 
\centering 
\includegraphics[scale=0.75]{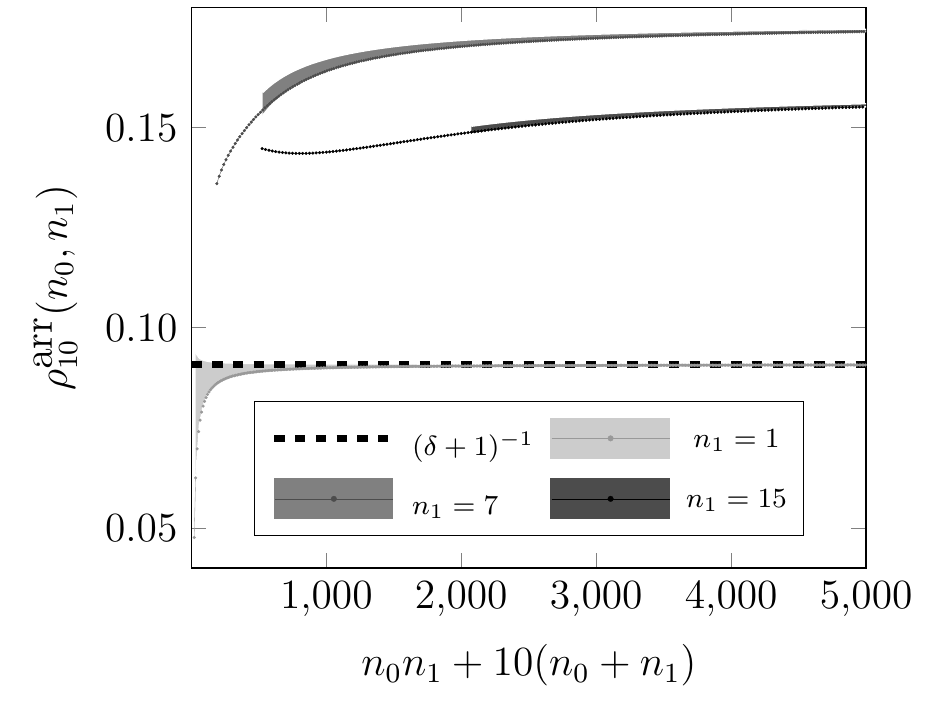} 
 \caption{The bit density of some example arrays as a function of area with $n_1$ fixed, parameterized by $n_0$ (with $n_0 \ge n_1$), with diode area factor $\delta = 10$.  The bounds from Corollary~\ref{cor:diodeDensityBounds} are included for the relevant values of $n_0$.} \label{fig:rhoVsA} 
\end{figure}

In Figure~\ref{fig:rhoVsA} we plot density against area for fixed values of the number of row wires $n_1$ and the number of column wires $n_0 \ge n_1$.  We plot information density for select values of $n_1$ and $\delta = 10$.  For $\delta = 10$, the information density as $n_0\to\infty$ becomes larger as we increase $n_1$ up to $n_1 = 7$ but decreases again and will ultimately approach zero as $n_1$ becomes large.

\subsection{Peripheral Circuitry for Single-Layer Devices}

In practice, the array area $A^{\textnormal{arr}}_\delta(n_0, n_1)$ is not the area of the entire device.  In this section, we take the peripheral circuitry needed for array element selection into consideration.  Suppose we tile a number of $n_0\times n_1$ devices into a rectilinear layout of $N_0$ rows and $N_1$ columns of memristive arrays.  To select a wire on each edge of the device, we will need demultiplexers of sizes $\log N_0n_0$-to-$N_0n_0$ and $\log N_1n_1$-to-$N_1n_1$.  Since a $\log n$-to-$n$ multiplexer has $\log n$ selection lines and $n$ output lines, each multiplexer will have an area that scales as $n\log n$ (see e.g.,~\cite[Sec. 8.3]{maini2007}).  This means that the overall device area is \begin{multline*}
 A_{\delta, \gamma}(N_0, N_1, n_0, n_1) := N_0N_1[n_0n_1 + \delta(n_0 + n_1)] +\\
\gamma(N_0n_0\log N_0n_0 + N_1n_1\log N_1n_1)
\end{multline*} where $\gamma$ is a positive constant.  This constant captures the scaling of the transistors in the demultiplexers with respect to the size of the individual array elements.  We also define the device information density as\begin{displaymath}
\rho_{\delta, \gamma}(N_0, N_1, n_0, n_1) := \frac{N_0N_1B(n_0, n_1)}{A_{\delta, \gamma}(N_0, N_1, n_0, n_1)}.
\end{displaymath}  We can then state the following theorem.

\begin{theorem}
If $n_0\to \infty$ and $N_1 \to \infty$ subject to the constraint that $N_1^{-1}\log n_0 \to 0$ and $n_0^{-1}\log N_1 \to 0$ with $N_0$ and $n_1$ fixed, then $\rho_{\delta, \gamma}(N_0, N_1, n_0, n_1) \to P^{\textnormal{arr}}_\delta(n_1)$.
\end{theorem}

\begin{IEEEproof}
Using Corollary~\ref{corr:bitBounds} we have, for sufficiently large $n_0$, \begin{multline*}
\frac{N_0N_1n_0\log(n_1 + 1)}{A_{\delta, \gamma}(N_0, N_1, n_0, n_1)}
\le \rho_\delta^\gamma(N_0, N_1, n_0, n_1) \le\\
 \frac{N_0N_1(n_0 + 1)\log(n_1 + 1)}{A_{\delta, \gamma}(N_0, N_1, n_0, n_1)}.
\end{multline*}  As $n_0 \to \infty$ and $y \to \infty$ this simplifies to \begin{multline*}
\frac{\log(n_1 + 1)}{n_1 + \delta + \gamma(\log n_0)/y + \gamma (n_1/x) (\log y)/n_0} \le\\
\rho_\delta^\gamma(N_0, N_1, n_0, n_1) \le\\
\frac{n_0 + 1}{n_0}\frac{\log(n_1 + 1)}{n_1 + \delta + \gamma(\log n_0)/y + \gamma (n_1/x) (\log y)/n_0}.
\end{multline*}  Since $(\log n_0)/y \to 0$ and $(\log y)/n_0 \to 0$, this means that $\rho_{\delta, \gamma}(N_0, N_1, n_0, n_1) \to \log(n_1 + 1)/(n_1 + \delta) = P^{\textnormal{arr}}_\delta(n_1)$, as $n, y_0 \to \infty$ as desired.
\end{IEEEproof}

Therefore, provided that the number of wires in the ``long'' dimension of the individual arrays and the number of arrays tiled along the ``short'' dimension grow such that neither is exponentially larger than the other, the peripheral circuitry from multiplexing has a negligible impact on the information density of large devices.  For devices whose overall aspect ratio is not extreme, this is a reasonable condition.  Consider for instance a device with constant positive finite aspect ratio $\alpha$, such that $N_0n_0 = \alpha N_1 n_1$.  This means that $(\log n_0)/N_1 = (\log(\alpha n_1 N_1/N_0))/N_1$ which approaches 0 as $N_1 \to \infty$, as desired.  By symmetry, $(\log N_1)/n_0$ also approaches 0.

\subsection{Multi-layer Density Optimization}

Having derived the capacity of multi-layer devices, we can also analyze their information density.  For simplicity, both of analysis and ultimately of circuit layout, let us assume that we have $\ell$ layers of memory cells of alternating dimensions, with $\ell$ an even number, so that our wire layer dimensions have the form $n\times m \times n \times m \times\cdots \times n$, so $n_{2i} = n$ and $n_{2i + 1} = m$.  We will refer to this as an $(m, n)_\delta^\ell$ device.  We state and prove multi-dimensional equivalents of Theorems~\ref{thm:stirlingBoundUpper} and~\ref{thm:stirlingBoundLower}.

\begin{theorem}\label{thm:multiUpper} 
If positive integers $m$ and $n$ satisfy \begin{equation}\label{eq:mBoundMulti}
m \ge \log_{1 + \frac1n}\frac1{27}n^3,
\end{equation} and $\ell$ is a positive even integer, then
 \begin{displaymath} 
T_\ell(m, n) \le \frac{[3^n(\beta_n - 1)]^{\ell/2 - 1}(n + 1)^\ell}{(2n + 1)^2} (2n + 1)^{\ell(m + 1)/2}
  \end{displaymath} 
  where $\beta_n := \left(\frac{0.792(n + 1)}{\ln(n + 2)}\right)^{n + 1}$.
\end{theorem}

\begin{IEEEproof}
See Appendix~\ref{app:bigProof}.
\end{IEEEproof}

\begin{theorem}\label{thm:multiLower} 
For integers $m, n > 0$,
\begin{displaymath} 
(2n + 1)^{\ell m/2} \le T_\ell(m, n). 
\end{displaymath} 
\end{theorem}

\begin{IEEEproof} 
Consider our device as a stack of $\frac\ell2$ two-layer $n\times m\times n$ subdevices.  We may do this if we respect the partitioning of the layers of $n$ wires shared between one device and the next.  The $i$th subdevice has array state matrix $\A = [(\A^{2i - 1})^\top|\A^{2i}]$, which is an $m\times 2n$ matrix.  If in each of these subdevices we connect each of the $m$ wires to at most one of the other $2n$ wires we will induce no sneak paths, since this is equivalent to having at most one 1 in each row of $\A$.  We also respect the partitioning, since each of the $2n$ wires is in a different partition (or disconnected).  There are $(2n + 1)^m$ ways of making these connections in each subdevice, for a total of $(2n + 1)^{\ell m/2}$ possible patterns in the device.
\end{IEEEproof}

Now suppose we isolate and tile our devices as before, introducing diodes of size $\delta$ on all wires connecting the tiled devices.  This means we add $\frac\ell2(m + n)$ diodes per tiled device.  Taking logarithms in Theorems~\ref{thm:multiUpper} and~\ref{thm:multiLower} and dividing by the array area, $A_\delta^{\ell, \textnormal{arr}}(m, n) := mn + \frac\ell2 \delta(m + n)$ gives the following corollaries.

\begin{corollary} 
The number of bits $B_\ell(m, n) := \log T_\ell(m, n)$ that can be stored in an $n\times m \times n \times\cdots\times n$ array with an even number of layers $\ell$ and dimensions satisfying~(\ref{eq:mBoundMulti}) satisfies \begin{multline*} 
\frac{\ell}2 m\log(2n + 1) \le B_\ell(m, n) \le \\ \log\frac{[3^n(\beta_n - 1)]^{\ell/2 - 1}(n + 1)^\ell}{(2n + 1)^2} + \frac{\ell}2(m + 1)\log(2n + 1).
\end{multline*} 
\end{corollary}

\begin{corollary} 
The array bit density $\rho_\delta^{\ell, \textnormal{arr}}(m, n)$ of an $(m, n)_\delta^\ell$ device with dimensions satisfying~(\ref{eq:mBoundMulti}) and $\ell$ even satisfies \begin{multline*} 
\frac{\frac\ell2 m\log(2n + 1)}{mn + \frac\ell2 \delta(m + n)} \le\rho_\delta^{\ell, \textnormal{arr}}(m, n) \le\\
\frac{\log\frac{[3^n(\beta_n - 1)]^{\ell/2 - 1}(n + 1)^\ell}{(2n + 1)^2} + \frac{\ell}2(m + 1)\log(2n + 1)}{mn + \frac\ell2 \delta(m + n)}.
\end{multline*} 
\end{corollary}

Taking $m\to\infty$ gives the following result.

\begin{corollary} For $\ell$ even and $n$ fixed, letting $m \to\infty$, the array bit density of an $(m, n)_\delta^\ell$ device converges to $P_\delta^{\ell, \textnormal{arr}}(n) := (n + \frac\ell2 \delta)^{-1}\frac\ell2\log(2n + 1)$. 
\label{corr:rhoMulti}
\end{corollary}

This is the result when $\ell$ is even, and we have an array of dimensions $n\times m \times \cdots \times n$.  If $\ell$ remains even, but the device has dimensions $m \times n \times \dots \times m$, the asymptotic bit density is $(n + \frac\ell2 \delta)^{-1}[(\frac\ell2 - 2)\log(2n + 1) + 2\log(n + 1)]$, while if $\ell$ is odd, the asymptotic bit density is $(n + \frac12(\ell + 1) \delta)^{-1}[\lfloor\frac\ell2\rfloor\log(2n + 1) + \log(n + 1)]$.

Corollary~\ref{corr:rhoMulti} shows that we can write the multi-layer asymptotic bit density in terms of the single layer density for an $(m, n)_\delta^\ell$ device with $\ell$ even.  In particular, $P_\delta^{\ell, \textnormal{arr}}(n) = \ell P_{\ell\delta}^{\textnormal{arr}}(2n)$.  This is reasonable, since we are effectively stacking $2n \times m$ devices in a total of $\ell$ layers, each of which still has diodes of size $\delta$.  This also means that we may use the same methods as in the single-layer case to find the value $n^*$ that maximizes the density.

The density $P_\delta^{\ell, \textnormal{arr}}(n)$ is an increasing function of $\ell$, but its derivative is decreasing.  This means that while adding more layers will always increase the information density, it will do so less and less efficiently as the layers are added (since the isolating diodes will still take up area on the lowest level in this model).  This means that a designer should choose $\ell$ by considering what fraction of the supremal density they consider acceptable, according to the following theorem.

\begin{theorem}
The density $P_\delta^{\ell, \textnormal{arr}}(n)$ satisfies \begin{equation}\label{eq:PBounds}
\frac{\log(2n + 1)}{n + \delta} \le P_\delta^{\ell, \textnormal{arr}}(n) < \frac{\log(2n + 1)}\delta
\end{equation} for fixed $n$ and $\delta$, and for any positive even value of $\ell$.  Furthermore, a density of at least $(1 - \epsilon)\delta^{-1}\log(2n + 1)$ is obtained by any choice of $\ell$ satisfying $\ell \ge 2(\epsilon^{-1} - 1)n\delta^{-1}$ when $0 < \epsilon \le 1$.
\end{theorem}

\begin{IEEEproof}
Since $P_\delta^{\ell, \textnormal{arr}}(n)$ is an increasing function of $\ell$, we may substitute $\ell = 2$ to obtain the lower bound in (\ref{eq:PBounds}) and take $\ell \to \infty$ to obtain the upper bound.  Solving $P_\delta^{\ell, \textnormal{arr}}(n) \le (1 - \epsilon)\frac{\log(2n + 1)}\delta$ for $\ell$ completes the proof.
\end{IEEEproof}

Note that the device area must be at least $A_{\delta, \gamma}^\ell(1, 1, m, n)$  to fit at least one array of size $m\times n$, plus the $\frac\ell2(m + n)$ diodes and the peripheral circuitry.  In practice, when choosing array dimensions and numbers of layers, there may be further concerns about power dissipation or the vertical size of the device, which are beyond the scope of the present model.  Note that if it is possible to stack the diodes along with the memristors, the density increases without bound as more layers are added in the limit of large $m$ and $N$.

\begin{figure}[!t] 
\centering 
\includegraphics[scale=0.75]{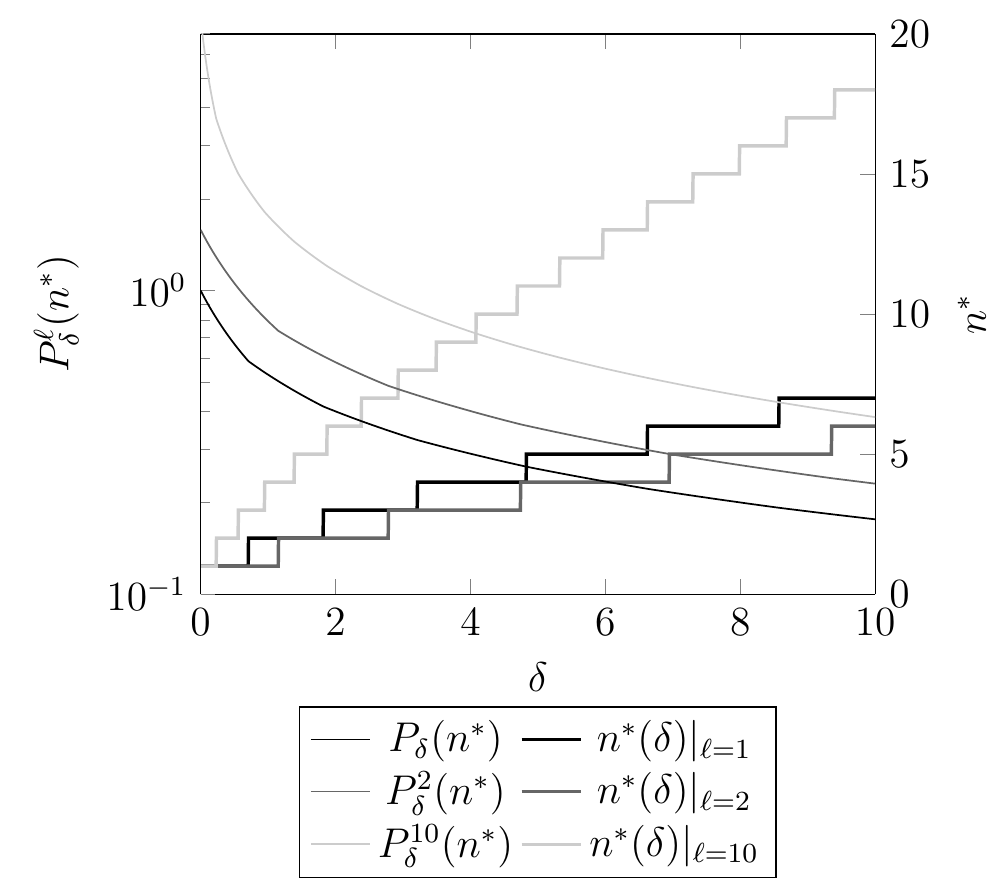}
 
\caption{Optimal array dimension $n^*$ versus diode area factor $\delta$ in the $m\to\infty$ regime (left axis) and the resulting bit density for this choice of $n$, for $\ell \in \{1, 2, 10\}$.} \label{fig:nStarMulti} 
\end{figure}

We plot the multi-layer equivalent of Figure~\ref{fig:nStar} in Figure~\ref{fig:nStarMulti}.  This plot includes the single-layer expression, as well as the $\ell = 2$ and $\ell = 10$ cases.  Note that $n^*$ is smaller for $\ell = 2$ than for $\ell = 1$, which is a result of the differences mentioned above between odd and even $\ell$s.

We can make similar arguments regarding peripheral circuitry in the multi-layer case to the single-layer case.  Define the device area \begin{multline*}
 A_{\delta, \gamma}^\ell(M, N, m, n) := MN[mn + \frac\ell2 \delta(m + n)] +\\
\gamma[M\frac\ell2m\log M\frac\ell2m + N\frac\ell2n\log N\frac\ell2n]
\end{multline*} and the device information density \begin{displaymath}
\rho_{\delta, \gamma}^{\ell, \textnormal{arr}}(M, N, m, n) := MNB_\ell(m, n)/A_{\delta, \gamma}^\ell(M, N, m, n)
\end{displaymath} where $M$, $N$, and $\gamma$ are defined as in the single-layer case.  We can then state the following theorem.

\begin{theorem}
If $m\to \infty$ and $N \to \infty$ such that $(\log m)/N \to 0$ and $(\log N)/m \to 0$ with $M$ and $n$ fixed, then $\rho_{\delta, \gamma}^{\ell, \textnormal{arr}}(M, N, m, n) \to P^{\ell, \textnormal{arr}}_\delta(n)$.
\end{theorem}

The proof is very similar to that of the single-layer case, and is omitted.

Note that taking a limit gives us an asymptotic expression for $T_\ell(m, n)$ for fixed $n$ and $m \to \infty$.

\begin{theorem}\label{thm:TLimitMulti}
For any fixed $n$ and even $\ell$, we have \begin{displaymath}
\lim_{m\to\infty}\frac{T_\ell(m, n)}{(2n + 1)^{\ell m/2}} = 1
\end{displaymath}
and this limit is approached from above.
\end{theorem}
\begin{IEEEproof}
The proof of this theorem is largely similar to that of Theorem~\ref{thm:TLimitSingle}, and is omitted.  The relevant observation is that by an analogous series of substitutions for the Stirling numbers of the form $\stirling{m + 1}{k_{2i + 1} + 1}$, $T_\ell(m, n) = (2n + 1)^{m\ell/2} + \mathcal{O}(2n)^{m\ell/2}$.
\end{IEEEproof}

\section{Encoding Schemes}

Now that we have seen how the information capacity of a resistive array scales asymptotically, we propose encoding and decoding schemes for storing and retrieving data that achieve this asymptotic capacity.  Our encoding and decoding schemes will make use of the following functions.

\begin{definition}[Binary conversion function]
Given a binary vector $(x_0, \dots, x_{k - 1}) \in \{0, 1\}^k$, define the binary conversion function $\phi_k: \{0, 1\}^k \mapsto \{0, 1, \dots, 2^k - 1\}$ by \begin{displaymath}
\phi_k(x_0, \dots, x_{k - 1})  = \sum_{i = 0}^{k - 1}x_i2^i
\end{displaymath} the value of the vector considered as a binary integer.
\end{definition}

Note that $\phi_k$ has a unique inverse.

\begin{definition}[High-bit set]\label{def:highBit}
The $i$th high-bit set of $k$-bit numbers is \begin{displaymath}
\phantom{,}\Delta_k(i) = \{j|\phi_k^{-1}(j)_i = 1, 0 \le j < 2^k\},
\end{displaymath} the set of all numbers that can be represented with $k$ bits, and whose $i$th bit is 1 in this representation.
\end{definition}

\subsection{Single-layer One-Hot Encoding Scheme}

To study questions of density, we need to consider a particular coding scheme.  The ``at-most-one-hot'' scheme that we will study is a simplification of the scheme in~\cite{cassuto2016}.  In this scheme we encode $n_0\log(n_1 + 1)$ bits into an $n_0\times n_1$ array, where $n_1 + 1$ is chosen to be a power of 2.  We will encode into a single device of this size.  In practice, many such devices will be tiled, as in Figure~\ref{fig:edgeDiode}, but we will encode and decode into each in isolation.

The data we will encode is $\{r_0, r_1, \dots, r_{n_0 - 1}\}$, where $r_i \in \{0, 1\}^{\log(n_1 + 1)}$ for each $i$ (this contains $n_0\log(n_1 + 1)$ bits of information).  Label the row nodes (wires) $R_0$ to $R_{n_0 - 1}$ and the column nodes $C_0$ to $C_{n_1 - 1}$.  We then encode the data into an $n_0\times n_1$ array as follows.
\begin{enumerate}
\item Set all array elements to the high resistance state.
\item Set the device at $(R_i, C_{\phi_{\log(n_1 + 1)}(r_i)})$ to the low resistance state if $\phi_{\log(n_1 + 1) + 1}(r_i) < n_1$ for each $i$ from 0 to $n_0 - 1$.
\end{enumerate}

That is, row $i$ contains a either a low resistance whose index corresponds to $r_i$ considered as a binary number, as in pulse-position modulation, or it contains only high resistances (when $\phi_{\log(n_1 + 1) + 1}(r_i) = n_1$).  As mentioned in the proof of Theorem~\ref{thm:stirlingBoundLower}, having at most one 1 in each row of the array state matrix guarantees that there are no sneak paths.

We decode as follows.  To find $r_i^j$, the $j$th bit of $r_i$, we make the measurement \begin{displaymath}
r_i^j = \mathcal{M}_G(\{R_i\}, \{C_p|p \in \Delta_{\log(n_1 + 1)}(j)\}).
\end{displaymath}  This gives us one bit per measurement.

This is a simpler version of the scheme proposed in~\cite{cassuto2016}.  The full scheme is described in Appendix~\ref{app:existingEncodingScheme}.  Our simplification is motivated by the fact that while the existing scheme achieves capacity asymptotically for arbitrary aspect ratios, it involves more complicated encoding and decoding, and does not allow for unique decoding of all data patterns.

When writing, if the pattern to be overwritten is known a priori, overwriting involves only changing those elements corresponding to the ones in the old and new positions\footnote{The mechanism for changing a memristor's state varies with the technology, but typically involves applying a large voltage to the device.  See~\cite{pan2014} for a discussion of some of the alternatives.}.  If the existing pattern is not known or the data is being initialized, the new pattern will be written in $k$ write operations corresponding to one row of $n_0$ array elements.

Note that writing is significantly more efficient when the data is known, since we need only change the old and new ``hot'' elements, rather than a whole row.  In contexts where write speed is a priority we may want first to read the current pattern and then to write, as this will take $k = \log (n_1 + 1)$ read and at most 2 write operations, instead of $2^k = n_1 + 1$ write operations.  The choice of writing strategy will depend on the specifics of the array size, read and write times, and the relative wear on the devices of reads and writes.  We leave this as a subject for future work.

\subsection{Multi-layer Encoding Scheme}

\subsubsection{Setup}

The exact capacity and density converge asymptotically to those of the at-most-one-hot encoding scheme for layered $m \times 2n$ devices as described above.  It is therefore natural to use such an encoding scheme for our device.

In particular, assume we have an $\ell$-layer device, with $\ell$ even.  For ease of encoding and decoding, we will consider a scheme that stores an integer number of bits.  However, since $2n + 1$ cannot be a power of 2, we cannot achieve the $\frac12 m\ell\log(2n + 1)$ limit exactly.  We can, however store $\frac\ell2 m\log(2n)$ bits, as will be described below.  The ratio between this value and the asymptotic value $\frac\ell2 m\log(2n + 1)$ is smallest at $n = 1$, where it is approximately equal to 0.63, and rapidly increases to an asymptotic value of 1 as $n$ increases (reaching 0.9 at $n \approx 2.5$, for instance), so this is not a large loss.

If we take $2n = 2^N$ for some integer $N$, the choice of $N$ which maximizes $P_\delta^{\ell,\textnormal{arr}}(n)$, which we will call $N^*$, will depend on $\delta$.  For an $(n, m)_\delta^\ell$ device we can store $\frac\ell2 mN$ bits per tiled $n \times m \times \dots \times n$ device, with an asymptotic density of $\widetilde P_\delta^{\ell,\textnormal{arr}}(N) := \ell N/(2^N + \ell\delta)$.  The following theorem will help develop a relationship between $N^*$ and~$\delta$ by finding values of $N$ where $\widetilde P$ is increasing.

\begin{theorem}
If $\ell$ is even, $\widetilde P_\delta^{\ell,\textnormal{arr}}(N)$ is increasing in $N$ when $\ell\delta \ge (N - 1)2^N$.
\end{theorem}

The proof is the result of a simple substitution into the expression for $\widetilde P$, and is omitted.  From this theorem we obtain the following corollary.

\begin{corollary}\label{corr:NChoice}
If integer $N$ has the largest value such that $\ell\delta \ge (N - 1)2^N$, with $\delta$ a fixed positive number and $\ell$ a fixed even number, then $N$ maximizes $\widetilde P_\delta^{\ell,\textnormal{arr}}$.
\end{corollary}

Optimal choices of $N$ over a range of values of $\ell\delta$ are given in Table~\ref{tab:NChoice}.

\begin{table}[!t]
\caption{Some values of $N^*$, and corresponding values of the lower bound on the range of $\ell\delta$ for which $N^*$ gives optimal density.}
\begin{displaymath}
\begin{IEEEeqnarraybox}[
\IEEEeqnarraystrutmode\IEEEeqnarraystrutsizeadd{3pt}{1pt}
]{l;v;c'c'c'c'c'c}
N^* && 1 & 2 & 3 & 4 & 5 & 6\\
\hline
\ell\delta && 0 & 4 & 16 & 48 & 128 & 320
\end{IEEEeqnarraybox}
\end{displaymath}
\label{tab:NChoice}
\end{table}

\subsubsection{Encoding scheme} 

We can now describe the encoding scheme for a device with even $\ell$ and dimensions $n \times m \times \cdots \times n$.  Similar schemes can be derived for odd $\ell$ or dimensions $m \times n \times \cdots \times m$.

We can think of our device as $\frac\ell2$ two-layer subdevices of size $n\times m \times n$.  Assign each of these subdevices a unique index $i$ with $0 \le i < \ell/2$.

The data we will encode is $\{r_{i, j}\}$, where $0 \le i < \ell/2$ and $0 \le j < m$, with each $r_{i,j} \in \{0, 1\}^{\log(2n)}$, for a total of $\frac\ell2 m\log(2n)$ bits.  We encode as follows.

\begin{enumerate}
\item Set all array elements to the high-resistance state.

\item Consider subdevice $i$.  Label the $m$ row wires from $R_0$ to $R_{m - 1}$ and the $2n$ column wires from $C_0$ to $C_{2n - 1}$.  For $j = 0, \dots, m - 1$, set the device at $(R_j, C_{\phi_{\log(2n)}(r_{i, j})})$ to the low-resistance state.

\item Repeat for all values of $i$.
\end{enumerate}

To decode the bit in $r_{i, j}^u$, the $u$th bit of $r_{i, j}$, perform the following operations.
\begin{enumerate}
\item Consider subdevice $i$.  Label the $m$ row wires from $R_0$ to $R_{m - 1}$ and the $2n$ column wires from $C_0$ to $C_{2n - 1}$ in the same order as when encoding.

\item Evaluate the measurement \begin{displaymath}
\mathcal{M}_G(\{R_j\}, \{C_p|p \in \Delta_{\log2n}(u)\})
\end{displaymath} with $\Delta$ defined as in Definition~\ref{def:highBit}, and $G$ the connectedness graph corresponding to the device. This is the desired bit.
\end{enumerate}

\begin{figure}[!t] 
\centering 
\includegraphics[scale=0.75]{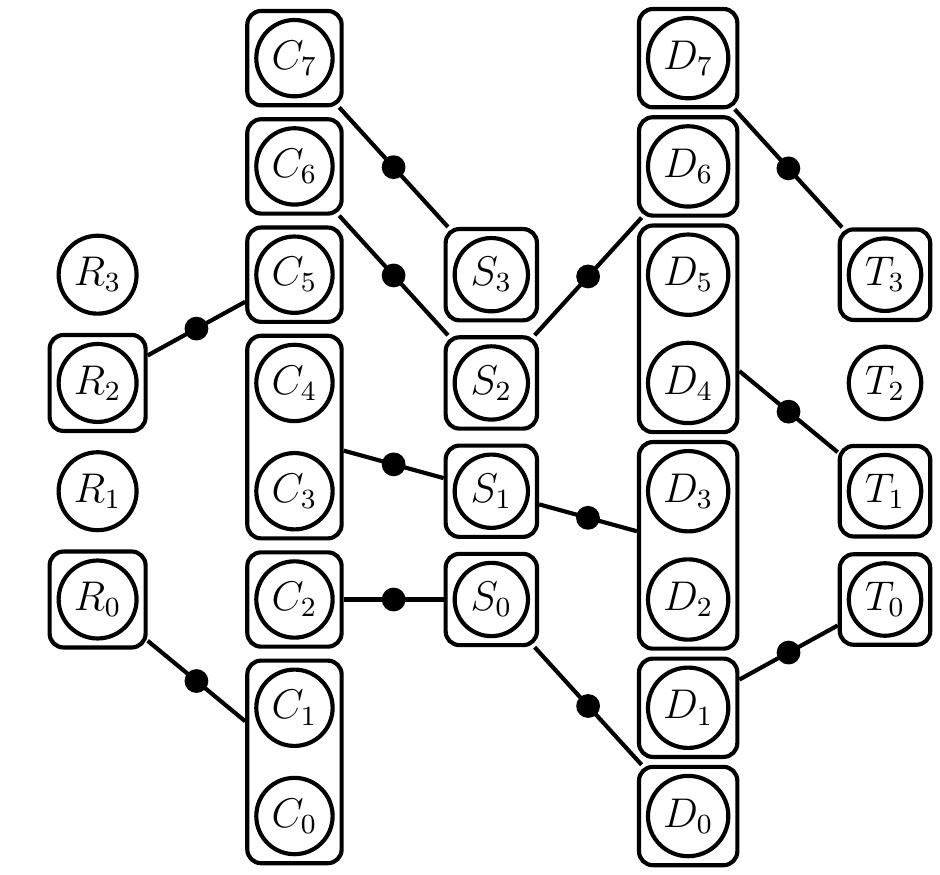}
 
\caption{A dot-graph for a sample array using one-hot encoding, with $m = 8$, $n = 4$, and $\ell = 4$.} \label{fig:dotGraphOneHot} 
\end{figure}

This decoding scheme requires one measurement per bit extracted.  An example dot-graph for a device using this scheme with $m = 8$, $n = 4$, and $\ell = 4$ is shown in Figure~\ref{fig:dotGraphOneHot}.  Note that each wire in the layers of $m$ wires is attached to exactly one wire in the adjacent layers of $n$ wires.

\section{Conclusions and Future Work}

We have explored the capacity of multi-layer resistive memory arrays, both exactly and asymptotically.  We have proposed information density as a useful figure of merit for characterizing coding techniques for resistive memory arrays of given dimension.

We have also shown that if diodes that contribute to the circuit area are used, one can achieve a higher information area density through a judicious choice of the placement of diodes, rather than by isolating every memory element.   Our derivations have shown that a simple encoding scheme achieves the maximum density asymptotically when using an array that consists of several smaller isolated subarrays.

Our model does not take into account power consumption considerations, which are useful when choosing between the varieties of isolation devices~\cite{shi2020}.  Power consumption is also relevant to some approaches that isolate smaller sections of an array by grounding some input and output lines.  This increases the array density, but also increases the power consumption~\cite{cassuto2016}.  An even more sophisticated model might model non-idealities in the system such as finite conductances and variations in array devices and sources.

Some technologies allow for diodes to be placed directly on top of memristors without an area cost, or include memristors with inherent selective properties (i.e., they act like diodes).  When this is possible, it is not necessary to tile isolated circuits as described here.  However, this may affect the power consumption or cost of the device.  More investigation is needed of the parameters that determine when the architecture discussed in this paper is preferable to the one with individual selection devices.

\appendices

\section{Proof of Theorem~\ref{thm:stirlingBoundUpper}}
\label{app:singleProof}

We introduce the following lemmas concerning bounds on sums of log-concave sequences, to help us to bound $T_1(n_0, n_1)$.  The following standard definition will be useful to this analysis (see, e.g., \cite{liu2007}).  

\begin{definition}[Log-concave sequence]
A sequence $f(k)$ is called logarithmically concave (or ``log-concave") if it satisfies $f^2(k) \ge f(k - 1)f(k + 1)$ for all integers $k \ge1$. 
\end{definition}

We will use this definition in the following lemma.

\begin{lemma} \label{lem:mBoundBasic} 
If $f$ is a positive log-concave sequence defined over integers between 0 and $n$, where $n$ is a non-negative integer such that $f(n) > f(n - 1)$, then \begin{displaymath} (n + 1)f(n) \ge\sum_{i = 0}^n f(i). \end{displaymath} 
\end{lemma}
\begin{IEEEproof} 
Since the summand $f(i)$ is log-concave, it has exactly one global maximum.  Since $f(n) > f(n - 1)$, the maximum must be at $i = n$, so we may bound each term by the $i = n$ term, which gives the desired result.  
\end{IEEEproof}

Note that this is usually not a tight bound, since we are bounding each term by the maximum.  We will more often use this lemma in the following form.

\begin{lemma}\label{lem:mBoundFG} 
If $f$ and $g$ are positive log-concave sequences defined over integers between 0 and $n$, with $f(n) > f(n - 1)$, and there exists an integer $m$ such that \begin{equation}\label{eq:mCond} 
m \ge\log_{\frac{f(n)}{f(n - 1)}}\frac{g(n - 1)}{g(n)} 
\end{equation}then \begin{equation}\label{eq:FGBound}
(n + 1)g(n)f(n)^m \ge\sum_{i = 0}^ng(i)f(i)^m. 
\end{equation} 
\end{lemma}
\begin{IEEEproof} 
If $f$ and $g$ are log-concave, and $m$ is a positive constant, then $gf^m$ is also log-concave, by application of the definition.  By Lemma~\ref{lem:mBoundBasic}, (\ref{eq:FGBound}) holds if \begin{displaymath}
g(n)f(n)^m \ge g(n - 1)f(n - 1)^m
\end{displaymath} which is equivalent to~(\ref{eq:mCond}). 
\end{IEEEproof}

The following lemma is useful for deriving bounds on information density for devices with extreme aspect ratios.

\begin{lemma} 
For integers $n$ and $k$ with $n > k$, $\stirling{n}{k} \le k^n/k!$. \label{lem:stirlingUpper} 
\end{lemma} 
\begin{IEEEproof} 
The expression $k^n/k!$ is the number of mappings from $[n] :=\{1, 2, \dots, n\}$ to $[k] :=\{1, 2, \dots, k\}$, where relabelings of the range correspond to the same mapping.  The partitions of $[n]$ into $k$ parts correspond to a subset of size $\stirling{n}{k}$ of these mappings, specifically those mappings whose image is the codomain $[k]$.\footnote{For $n \gg k$, with $k$ fixed, $k^n/k!$ is a good approximation for $\stirling{n}{k}$ since the probability that random elements of $[k]$ are mapped to by some element in $[n]$ is very high, if mappings are made uniformly at random.} 
\end{IEEEproof}

We may now restate and prove Theorem~\ref{thm:stirlingBoundUpper}.

\begin{reptheorem}{\ref{thm:stirlingBoundUpper}} 
If positive integers $n_0$ and $n_1$ satisfy\begin{equation} n_0 \ge\log_{(1 + \frac1{n_1})}\frac{n_1(n_1 + 1) }2 \label{eq:mBound} 
\end{equation} then $T_1(n_0, n_1) \le(n_1 + 1)^{n_0 + 1}$.
\end{reptheorem} 
\begin{IEEEproof} 
Use Lemma~\ref{lem:stirlingUpper} to write 
	\begin{IEEEeqnarray*}{rCl}
	T_1(n_0, n_1) &=& \sum_{k = 0}^{n_1}\stirling{n_1 + 1}{k + 1}\stirling{n_0 + 1}{k + 1}k!\\
	 &\le& \sum_{k = 0}^{n_1}\stirling{n_1 + 1}{k + 1}(k + 1)^{n_0}. 
	\end{IEEEeqnarray*}
Since $\stirling{n_1 + 1}{k + 1}$ is log-convex in $k$~\cite{lieb1968}, we may take $g(k) = \stirling{n_1 + 1}{k + 1}$ and $f(k) = k + 1$ in Lemma~\ref{lem:mBoundFG}, which we may apply when~(\ref{eq:mBound}) holds.   This gives $T_1(n_0, n_1) \le(n_1 + 1)^{n_0 + 1}$ as desired.
\end{IEEEproof}

\section{Proof of Theorem~\ref{thm:multiUpper}}
\label{app:bigProof}

\begin{reptheorem}{\ref{thm:multiUpper}} 
If positive integers $m$ and $n$ satisfy \begin{equation}\label{eq:mBoundRep}
m \ge \log_{1 + \frac1n}\frac{n^3}{27},
\end{equation} and $\ell$ is a positive even integer, then
 \begin{displaymath} 
T_\ell(m, n) \le \frac{[3^n(\beta_n - 1)]^{\ell/2 - 1}(n + 1)^\ell}{(2n + 1)^2} (2n + 1)^{\ell(m + 1)/2}
  \end{displaymath} 
  where $\beta_n := \left(\frac{0.792(n + 1)}{\ln(n + 2)}\right)^{n + 1}$.
\end{reptheorem}

\begin{IEEEproof} 
By Theorem~\ref{thm:TMulti}
	\begin{multline}\label{eq:TMulti}
T_\ell(m, n) = \sum_{s_1 = 0}^n\cdots\sum_{s_\ell = 0}^n\Bigg[\prod_{i = 0}^{\ell/2 - 1}s_{2i + 1}!s_{2i + 2}!\\
\sum_{k_{2i + 1} = \max(s_{2i + 1}, s_{2i + 2})}^{s_{2i + 1} + s_{2i + 2}}\widetilde Q_m(k_{2i + 1}, s_{2i + 1}, s_{2i + 2})\Bigg]\\
\left[\prod_{i = 0}^{\ell/2}\sum_{k_{2i} = \max(s_{2i}, s_{2i + 1})}^{\min(n, s_{2i} + s_{2i + 1})}\widetilde Q_n(k_{2i}, s_{2i}, s_{2i + 1})\right]
\end{multline} where $s_0 = s_{\ell + 1} = 0$, and \begin{displaymath}
\widetilde Q_n(k, a, b) := \binom{k}{a + b - k, k - a, k - b}\stirling{n + 1}{k + 1}
\end{displaymath}
Use Lemma~\ref{lem:stirlingUpper} to bound $\stirling{m + 1}{k_i + 1}$ and define \begin{displaymath}
h_{s_0, s_1}(k) := (s_0 + s_1 - k)^+!(k - s_1)^+!(k - s_0)^+!,
\end{displaymath} where $x^+ := \max(0, x)$.  This means that \begin{displaymath}
\widetilde Q_m(k_{2i + 1}, s_{2i + 1}, s_{2i + 2}) \le \frac{(k_{2i + 1} + 1)^m}{h_{s_{2i + 1}, s_{2i + 1}}(k_{2i + 1})}.
\end{displaymath}
We apply Lemma~\ref{lem:mBoundFG} for each $i$, with $g_i(k_{2i + 1}) = k_{2i + 1} + 1$ and $f_i(k_{2i + 1}) = [h_{s_{2i + 1}, s_{2i + 1}}(k_{2i + 1})]^{-1}$.   The tightest of the resulting bounds in~(\ref{eq:FGBound}) is~(\ref{eq:mBoundRep}).  This gives
\begin{IEEEeqnarray}{rCl}
\IEEEeqnarraymulticol{3}{l}{
	\sum_{k_{2i + 1} = \max(s_{2i + 1}, s_{2i + 2})}^{s_{2i + 1} + s_{2i + 2}}\widetilde Q_m(k_{2i + 1}, s_{2i + 1}, s_{2i + 2})
}\nonumber\\
\qquad\qquad& \le &\sum_{k_{2i + 1} = 0}^{s_{2i + 1} + s_{2i + 2}} \frac{(k_{2i + 1} + 1)^m}{h_{s_{2i + 1}, s_{2i + 1}}(k_{2i + 1})}  \nonumber\\
& \le & \frac{(s_{2i + 1} + s_{2i + 2} + 1)^{m + 1}}{s_{2i + 1}!s_{2i + 2}!}.\label{eq:QmBound}
\end{IEEEeqnarray}

Substituting~(\ref{eq:QmBound}) into~(\ref{eq:TMulti}) gives
\begin{multline}\label{eq:TMultiIntermediate}
T_\ell(m, n) \le \sum_{s_1 = 0}^n\cdots\sum_{s_\ell = 0}^n \left[\prod_{i = 0}^{\ell/2 - 1}(s_{2i + 1} + s_{2i + 2} + 1)^{m + 1}\right]\\
\Bigg[\prod_{i = 0}^{\ell/2}\sum_{k_{2i} = \max(s_{2i}, s_{2i + 1})}^{\min(n, s_{2i} + s_{2i + 1})}\widetilde Q_n(k_{2i}, s_{2i}, s_{2i + 1})\Bigg].
\end{multline}

Since the sums over $k_0$ and $k_\ell$ each have only one term, $\sum_{k_0}\widetilde Q_n(k_0, s_0, s_1) \le \stirling{n + 1}{s_1 + 1}$ and $\sum_{k_\ell}\widetilde Q_n(k_\ell, s_\ell, s_{\ell + 1}) \le \stirling{n + 1}{s_\ell + 1}$.  For the other sums over $k_{2i}$, \begin{IEEEeqnarray*}{rCl}
\IEEEeqnarraymulticol{3}{l}{
\sum_{k_{2i} = \max(s_{2i}, s_{2i + 1})}^{\min(n, s_{2i} + s_{2i + 1})}\widetilde Q_n(k_{2i}, s_{2i}, s_{2i + 1})
}\\
 \qquad\qquad&\le &\sum_{k_{2i} = \max(s_{2i}, s_{2i + 1})}^{\min(n, s_{2i} + s_{2i + 1})} 3^{k_{2i}}\stirling{n + 1}{k_{2i} + 1}\\
  &\le &3^n\sum_{k_{2i} = 0}^n \stirling{n + 1}{k_{2i} + 1}\\
  &=& 3^n(B_{n + 1} - 1)
\end{IEEEeqnarray*}
where the Bell number $B_n := \sum_{k = 0}^n\stirling{n}{k}$ and we have used $\sum_{a, b, c}\binom{x}{a, b, c} = 3^x$ to bound the trinomial coefficients. 

Substituting into~(\ref{eq:TMultiIntermediate}) gives 
\begin{multline} \label{eq:TMultiS}
T_\ell(m, n) \le [3^n(B_{n + 1} - 1)]^{\ell/2 - 1}\\
\left(\sum_{s_1 = 0}^n\sum_{s_2 = 0}^n \stirling{n + 1}{s_1 + 1}(s_1 + s_2 + 1)^{m + 1}\right)\\
\left(\sum_{s_{\ell - 1} = 0}^n\sum_{s_\ell = 0}^n \stirling{n + 1}{s_\ell + 1}(s_{\ell - 1} + s_\ell + 1)^{m + 1}\right)\\
\prod_{i = 1}^{\ell/2 - 2}\left(\sum_{s_{2i + 1} = 0}^n\sum_{s_{2i + 2} = 0}^n(s_{2i + 1} + s_{2i + 2} + 1)^{m + 1}\right)
\end{multline}
Applying Lemma~\ref{lem:mBoundFG} twice to each bracketed sum in~(\ref{eq:TMultiS}) gives \begin{displaymath}
T_\ell(m, n) \le \frac{[3^n(B_{n + 1} - 1)]^{\ell/2 - 1} (n + 1)^\ell}{(2n + 1)^2}(2n + 1)^{\ell(m + 1)/2}
\end{displaymath} where the bounds from~(\ref{eq:mCond}) are all dominated by the bound in~(\ref{eq:mBoundRep}).

 Using~\cite[Theorem 2.1]{berend2000}, we bound $B_{n + 1}$ by $\beta_n$, giving \begin{displaymath} 
T_\ell(m, n) \le \frac{[3^n(\beta_n - 1)]^{\ell/2 - 1}(n + 1)^\ell}{(2n + 1)^2} (2n + 1)^{\ell(m + 1)/2}
\end{displaymath} as desired.
\end{IEEEproof}

\section{Comments on an Existing Encoding Scheme}
\label{app:existingEncodingScheme}

In~\cite{cassuto2016}, the following single-layer encoding scheme is proposed.  Given a device of size $n_0\times n_1$ with array state matrix $\A$, select parameter $\lambda$ to be a power of 2 with $0 \le \lambda < n_1$.  The data is vectors $\{r_0, \dots, r_{n_0 - 1}, c_0, \dots, c_{n_1 - \lambda - 1}\}$, where $r_i, c_j \in \{0, 1\}^{\log\lambda}$ for each $i$ and $j$.  Let $\A_{:, i}$ denote the $i$th column of $\A$.  Then we encode as follows:

\begin{enumerate}
\item Set all $\A_{ij}$ to 0.
\item For $i = 0, \dots, n_0 - 1$, set $\A_{i, \phi_{\log\lambda}(r_i)} = 1$.
\item For $j = 0, \dots, n_1 - \lambda - 1$, set $\A_{:, j + L} = \A_{:, \phi_{\log\lambda}(c_j)}$.
\end{enumerate}

This scheme encodes $(n_0 + n_1 - \lambda)\log\lambda$ bits.  If we choose $\lambda$ to be the power of 2 closest to $(n_0 + n_1)/\log(n_0 + n_1)$, this number of bits approaches capacity asymptotically~\cite{cassuto2016}.  This encoding scheme can result in collisions, however, as will be seen in the following example.  In addition, multiple measurements are required to decode each of the $c_i$s, which complicates decoding.

\begin{table}[!t]
\caption{Data used in Example~\ref{ex:collision}.}\label{tab:data}
\begin{displaymath}
\begin{IEEEeqnarraybox}[
\IEEEeqnarraystrutmode\IEEEeqnarraystrutsizeadd{3pt}{1pt}
]{c;v;c'c'c;v;c'c'c}
&& r_0 & r_1 & r_2 && c_0 & c_1 & c_2\\
\hline
\text{Data 0} && \{0, 0\} & \{0, 1\} & \{0, 0\} && \{0, 1\} & \{0, 0\} & \{1, 0\}\\
\text{Data 1} && \{0, 0\} & \{0, 1\} & \{0, 0\} && \{0, 1\} & \{0, 0\} & \{1, 1\}
\end{IEEEeqnarraybox}
\end{displaymath}
\end{table}

\begin{example}\label{ex:collision}
Consider the case where $n_0 = 3$, $n_1 = 7$, and $\lambda = 4$, and we want to encode the data from row ``Data 0'' of Table~\ref{tab:data}.  This data is encoded into the following array state matrix: \begin{equation}\label{eq:dataEncoding}
\left[\begin{tabular}{cccc|cccccc}
1 & 0 & 0 & 0 & 0 & 1  & 0\\
0 & 1 & 0 & 0 & 1 & 0  & 0\\
1 & 0 & 0 & 0 & 0 & 1  & 0\\
\end{tabular}\right]
\end{equation}

If, instead, we encode the distinct information in row ``Data~1'' of Table~\ref{tab:data}, the same array state matrix is obtained, so we cannot decode uniquely.
\end{example}

\section*{Acknowledgments}

The authors wish to thank the reviewers of this paper, whose insightful questions and comments have greatly improved the scope and depth of the material presented.


\begin{IEEEbiographynophoto}{Susanna~E.~Rumsey}
(Graduate Student Member, IEEE) was born in Toronto, ON, Canada in 1993.  She
received the B.A.Sc. degree with honours in engineering science (major in engineering
physics), and the M.Eng and M.A.Sc.~degrees in electrical and computer
engineering from the University of Toronto, Toronto, ON, Canada, in 2015, 2016,
and 2019 respectively.

Since 2019, she has been a Ph.D. student in electrical and computer engineering
at the University of Toronto, Toronto, ON, Canada.
\end{IEEEbiographynophoto}

\begin{IEEEbiographynophoto}{Stark C. Draper}
(Senior Member, IEEE) received the M.S. and Ph.D. degrees from the
Massachusetts Institute of Technology (MIT), and the B.S. and B.A. degrees in
electrical engineering and in history from Stanford University. He is a
Professor of Electrical and Computer Engineering with the University of Toronto
(UofT) and was an Associate Professor with the University of Wisconsin,
Madison. As a Research Scientist he has worked with the Mitsubishi Electric
Research Labs (MERL), Disney's Boston Research Lab, Arraycomm Inc., the C. S.
Draper Laboratory, and Ktaadn Inc. He completed Postdocs with the UofT and at
the University of California, Berkeley. His research interests include
information theory, optimization, error-correction coding, security, and the
application of tools and perspectives from these fields in communications,
computing, and learning.

Prof. Draper is a recipient of the NSERC Discovery Award, the NSF CAREER Award, the 2010 MERL President's Award, and teaching awards from the UofT, the University of Wisconsin, and MIT. He received an Intel Graduate Fellowship, Stanford's Frederick E. Terman Engineering Scholastic Award, and a U.S. State Department Fulbright Fellowship. He spent the 2019--2020 academic year on sabbatical at the Chinese University of Hong Kong, Shenzhen, and visiting the Canada-France-Hawaii Telescope (CFHT) in Hawai'i, USA.  He chairs the Machine Intelligence major at UofT, is a member of the IEEE Information Theory Society Board of Governors, and serves as the Faculty of Applied Science and Engineering representative on the UofT Governing Council.
\end{IEEEbiographynophoto}
\vfill
\newpage

\begin{IEEEbiographynophoto}{Frank R. Kschischang}
(Fellow, IEEE) received the B.A.Sc.\ degree (Hons.) from the University of
British Columbia, Vancouver, BC, Canada, in 1985,and the M.A.Sc.\ and Ph.D.
degrees from the University of Toronto, Toronto, ON, Canada, in 1988 and 1991,
respectively, all in electrical engineering. Since 1991, he has been a Faculty
Member in electrical and computer engineering with the University of Toronto,
where he currently holds the title of Distinguished Professor of digital
communication. His research interests are in the area of channel
coding techniques, applied to wireline, wireless, and optical communication
systems and networks.

Dr. Kschischang is a Fellow of the Engineering Institute of Canada, the Canadian Academy of Engineering, and the Royal Society of Canada. He has received several awards for teaching and research, including the 2010 Communications Society and Information Theory Society Joint Paper Award and the 2018 IEEE Information Theory Society Paper Award. He served as the General Co-Chair for the 2008 IEEE International Symposium on Information Theory. He served as the 2010 President of the IEEE Information Theory Society.  He received the Society's Aaron D. Wyner Distinguished Service Award in 2016. From 1997 to 2000, he served as an Associate Editor for Coding Theory for the \textsc{IEEE Transactions on Information Theory}, for which he served as the Editor-in-Chief from 2014 to 2016.
\end{IEEEbiographynophoto}
\end{document}